\documentclass[english,nolineno]{socg-lipics-v2021}

\usepackage{amsmath, graphicx, mathtools, url, enumitem, apxproof, siunitx}
\usepackage{tikz, pgfplots, filecontents, multirow}
\usepackage{algorithm, algpseudocode}
\usepackage{subcaption}
\usepackage[inkscapeformat=png]{svg}
\usepackage[labelformat=simple]{subcaption}
\usepackage{xcolor}

\hideLIPIcs

\renewcommand\thesubfigure{(\alph{subfigure})}

\usetikzlibrary{patterns, positioning, arrows.meta, shapes, positioning, fit, backgrounds, ext.paths.ortho, trees}

\algblockdefx{MRepeat}{EndRepeat}{\textbf{repeat}}{}
\algnotext{EndRepeat}

\algblock{ParWhile}{EndParWhile}
\algnewcommand\algorithmicparwhile{\textbf{par-while}}
\algnewcommand\algorithmicpardo{\textbf{do}}
\algnewcommand\algorithmicendparwhile{\textbf{end\ par-while}}
\algrenewtext{ParWhile}[1]{\algorithmicparwhile\ #1\ \algorithmicpardo}
\algrenewtext{EndParWhile}{\algorithmicendparwhile}

\newtheorem{thm}{Theorem}
\newtheorem{lem}{Lemma}
\newtheorem{rmk}{Remark}
\newtheorem{pro}{Problem}
\newtheorem{col}{Corollary}
\newtheorem{mydef}{Definition}
\newtheorem{eg}{Example}

\definecolor{OI1}{RGB}{230,159,0}
\definecolor{OI2}{RGB}{86,180,233}
\definecolor{OI3}{RGB}{0,158,115}
\definecolor{OI4}{RGB}{240,228,66}
\definecolor{OI5}{RGB}{204,121,167}
\definecolor{OI6}{RGB}{213,94,0}
\definecolor{lightergray}{RGB}{210, 210, 210}

\title{Finding a Fair Scoring Function for Top-$k$ Selection: From Hardness to Practice}

\author{Guangya Cai}{University of Minnesota, Twin Cities, MN, USA}{cai00171@umn.edu}{}{}

\authorrunning{G. Cai}

\Copyright{Guangya Cai}

\ccsdesc[500]{Information systems~Top-k retrieval in databases}

\keywords{Fairness, Top-$k$, Integration}

\supplement{\textit{Artifact Availability}: The source code, data, and/or other artifacts have been made available at \url{https://github.com/caiguangya/fair-topk}.}

\acknowledgements{The author would like to acknowledge the Minnesota Supercomputing Institute (MSI) at the University of Minnesota for providing computing facilities.}

\begin{document}

\maketitle

\begin{abstract}
    Selecting a subset of the $k$ ``best'' items from a dataset of $n$ items, based on a scoring function, is a key task in decision-making. Given the rise of automated decision-making software, it is important that the outcome of this process, called top-$k$ selection, is fair. Here we consider the problem of identifying a fair linear scoring function for top-$k$ selection. The function computes a score for each item as a weighted sum of its (numerical) attribute values, and must ensure that the selected subset includes adequate representation of a minority or historically disadvantaged group. Existing algorithms do not scale efficiently, particularly in higher dimensions. Our hardness analysis shows that in more than two dimensions, no algorithm is likely to achieve good scalability with respect to dataset size (i.e., a run time of $O(n\cdot \text{polylog}(n))$), and the computational complexity is likely to increase rapidly with dimensionality. However, the hardness results also provide key insights guiding algorithm design, leading to our two-pronged solution: (1) For small values of $k$, our hardness analysis reveals a gap in the hardness barrier. By addressing various engineering challenges, including achieving efficient parallelism (yielding up to two orders of magnitude speedup with 128 cores), we turn this potential of efficiency into an optimized algorithm delivering substantial practical performance gains. (2) For large values of $k$, where the hardness is robust, we employ a practically efficient algorithm which, despite being theoretically worse, achieves superior real-world performance. Experimental evaluations on real-world datasets then explore scenarios where worst-case behavior does not manifest, identifying areas critical to practical performance. Our solution achieves speedups of up to several orders of magnitude compared to the state of the art (SOTA), an efficiency made possible through a tight integration of hardness analysis, algorithm design, practical engineering, and empirical evaluation.
\end{abstract}

\section{Introduction}
Selecting a subset of items from a large dataset, based on certain criteria, is a critical task in decision-making. Typically, a score is computed for each item based on the item's attributes and then a subset consisting of the highest-scoring items is selected. For example, to decide whom to interview, a recruiter may evaluate multiple attributes (e.g., years of experiences and the relevance of previous jobs) of an applicant and assign each applicant an overall score reflecting their qualifications. Then a small subset is selected by score. To arrive at an overall score, a scoring function is necessary. Among the many different types of scoring functions, a linear-weighted function is often used. Here each numerical attribute is assigned a weight (or preference) to reflect its relative importance and a weighted sum of these attributes' values determines the overall score. The problem of selecting the $k$ highest-scoring items from a dataset of $n$ items ($k\leq n$) is called \emph{Top-$k$ Selection}.

Given the widespread usage of automated software for decision-making these days, there is an increased emphasis on ensuring that the results of tasks such as top-$k$ selection are fair. Among the different ways of quantifying fairness (e.g.,~\cite{friedler2021possibility,yang2017measuring,zehlike2022fairness}), a common one is based on ensuring proportional representation, which works as follows: In addition to numerical attributes, each item also has categorical attributes (gender, race, ethnicity, etc.). Typically, one or a combination of these attributes is regarded as a \textit{sensitive} attribute and one value of this sensitive attribute indicates a membership in a minority or historically disadvantaged group, often referred to as a \textit{protected} group. The goal then is to identify a top-$k$ subset in which the items of the protected group are in roughly the same proportion as they are in the entire dataset. The degree of fairness of a top-$k$ selection is measured by the difference between the proportion of protected group members in the selected subset and their proportion in the entire dataset.

In many applications, one often wants top-$k$ selection results that are both fair and high-quality. To ensure a fair top-$k$ selection that also satisfies other needs, proportional fairness constraints are commonly applied to the results. In many of the previous works (e.g., \cite{zehlike2017fa, celis2018ranking, yang2019balanced}), these fairness constraints are applied after the scores of items are determined, which might result in different selection criteria for different groups. However, an intervention in selection criteria after scores are determined might be subject to legal challenges (e.g., Ricci vs. DeStefano \cite{ricci}), so using a fair scoring function to begin with could be a better choice. Besides, one might only be interested in finding a fair scoring function in a subset of scoring functions. For example, to select the top-$k$ applicants to interview for a position, a recruiter might believe the relevance of the applicant's previous jobs is more important than the applicant's years of experiences, so they might assign a larger weight to the former attribute than to the latter while designing the scoring function. The other situation could be that one already has a scoring function that does not produce a fair top-$k$ selection, and wants to find a ``near-by'' scoring function that produces a fair one, that is, each weight of the attribute does not vary from its original value by a lot. Sometimes even knowing there does not exist a ``near-by'' one could be helpful since that might reveal some negligence in the process of designing the scoring function. Consider the following variation of the example from \cite{asudeh2019designing}:

\begin{eg}\label{eg:admission}
    A college admissions officer is developing a selection scheme for a pool of applicants, each possessing multiple relevant attributes. For simplicity, let us consider two of these attributes: high school GPA and SAT score (denoted as $g$ and $s$), assuming both of these attributes are properly normalized and standardized. Suppose the given fairness constraint is that at least $40\%$ women should be in the admitted class. The officer might believe that $g$ and $s$ should have an approximately equal weight, thus having the scoring function $f(c) = 0.5 \times g + 0.5 \times s$ to compute a score for each applicant $c$. Then, the (say) top-500 applicants are selected by score.
    
    However, the resulting top-500 applicants might have an insufficient number of women: say, 150 women in the top-500 whereas 200 is desired. This violation may be due to a gender disparity in the SAT score, namely women scored lower on average than men \cite{sat}. The officer might try to fix the scoring function first, by finding a fair scoring function that is close to the original one. By doing so, the choice of weights (approximately equal weighting) is still explainable. The weights of $g$ and $s$ can range between $0.45$ and $0.55$. By running an algorithm for finding a fair scoring function, the officer might successfully find a fair scoring function $f(c) = 0.55 \times g + 0.45 \times s$.
    
    However, if no such function within the allowable weight range exists, that is, reducing the weight of $s$ by 0.05 is still insufficient, the algorithm reports a failure. This outcome suggests that the officer's design principle (assigning approximately equal weights) may not be appropriate for the fairness consideration. Upon reflection, the officer may choose to abandon the equal-weight design choice. Since the fairness concern may justify a deviation from equal weighting, the officer may manually decrease the weight of $s$ and, with the aid of the algorithm for finding a fair scoring function, arrive at a new scoring function $f(c) = 0.6\times g + 0.4 \times s$ that meets the fairness constraint.
\end{eg}

There is work on finding a fair scoring function \cite{asudeh2019designing}, but known algorithms do not scale well with the size of the dataset (with the worst case run time being $\Omega(n^2)$). Even worse, their run times increase dramatically with dimensionality, which greatly limits their practical usage. Of course, the fairness constraints considered in \cite{asudeh2019designing} are general ones, which leaves room for improving the run time for a more specific fairness constraint. In this work, we consider the simplest possible proportional fairness constraint on top-$k$ selection, which does not enforce the constraint on the relative rankings of items in the top-$k$ selection result (as in the Example~\ref{eg:admission}). This setting is also relevant in practice. For example, the relative ranking of applicants for interview selection is often ignored as the hiring decision is often primarily made by their performance during the interview. So, for such a simple fairness constraint, do we have better algorithms that are efficient for large datasets in practice (preferably run in $O(n\cdot \text{polylog}(n))$ times), especially for higher dimensional data?

To answer this question, we present a thorough study of the problem of finding a fair linear scoring function for top-$k$ selection. Our main contributions are as follows:
\begin{itemize}
    \item \textbf{Hardness analysis.} We establish the first known lower bounds for the problem, showing that a universally efficient algorithms is unlikely to exist for any dimensions $d\geq 3$ and the computational complexity is likely to increase rapidly with dimensionality. In fixed (constant) dimensions, we prove a fine-grained complexity lower bound of $\Omega(n^{2-\delta})$ (for any constant $\delta > 0$) and a decision-tree lower bound of $\Omega(n^{d-1})$, both for $d\geq3$. In arbitrary dimensions, we prove the NP-hardness of the problem. Moreover, we further prove that these lower bounds are also applicable to any reasonable proportional fairness constraint and any $k = \beta n$, where $0<\beta<1$ is a constant, matching many real-world scenarios.
    \item \textbf{Practical algorithms.} Guided by the hardness results, we propose an efficient two-pronged solution: (1) For small values of $k$, we design a practical algorithm based on the $k$-level that exploits an structural insight revealed by our hardness analysis. By addressing multiple engineering challenges, including achieving efficient parallelism (which results in up to two orders of magnitude speedup with $128$ cores), the algorithm achieves substantial real-world performance gains. (2) For large values of $k$, where the worst-case hardness is robust, we employ a practically efficient algorithm based on mixed integer linear programming (MILP), which works efficiently for real-world datasets.
    \item \textbf{Experimental evaluation.} We conduct an extensive experimental evaluation on real-world datasets, demonstrating speedups of up to several orders of magnitude over the state-of-the-art method. In particular, our analysis of the experimental results explores scenarios in which worst-case behavior does not manifest, identifying areas critical to practical performance. Moreover, we discuss the choice between the two algorithms in our two-pronged solution based on the experimental results.
\end{itemize}
\textbf{The integration of these components is central.} While some of the results above may be of independent interest, their primary significance in this work lies in their interconnection. Our goal is to deliver a practically efficient solution, which is achieved through the integration of hardness analysis, algorithm design, practical engineering and empirical evaluation, with each component working together to solve the problem. (See Figure 1 for the roadmap of these interconnections.)

\begin{figure}[hbpt]
    \centering
    \resizebox{\textwidth}{!}{
    \begin{tikzpicture}[
  box/.style = {draw=#1!30!white!35!black, very thick, rectangle, fill=#1!30!white!40!lightergray, text width=80pt, align=center, minimum height=30pt},
  cluster/.style = {draw, thick, fill=#1!30, inner sep=10pt, rounded corners},
  subcluster/.style = {draw=#1!35!black, dashed, ultra thick, fill=none, inner sep=8pt, rounded corners},
  node distance=12pt,
  level distance=100pt
]

  \node[box=OI2] (thm4)  {NP-hardness\\(Theorem~\ref{thm:np})};
  \node[box=OI2, above=16pt of thm4] (cor2) {Constraint Relax.\\(Corollary~\ref{col:relaxed})};
  \node[box=OI2, above=of cor2] (thm2) {$k$-Generalization\\(Theorem~\ref{thm:k})};
  \node[box=OI2, above=26pt of thm2] (cor1) {Curse of Dim.\\(Corollary~\ref{col:hd})};
  \node[box=OI2, above=of cor1] (thm1) {3SUM-hardness\\(Theorem~\ref{thm:3sum})};
  \node[inner sep=0pt, above= 20pt of thm1] (hard) {\Large\textbf{Hardness Analysis}};
  \node[subcluster=OI2, fit=(cor1)(thm1)] {};
  \node[subcluster=OI2, fit=(thm2)(cor2)] {};
  \begin{scope}[on background layer]
      \node[cluster=OI2, fit=(hard)(thm4)(cor2)(thm2)(cor1)(thm1), anchor=center] (hardness_cluster) {};
  \end{scope}

\node[box=OI3, right=100pt of cor1] (smallk) {Small $k$\\Opportunity};
\node[box=OI3, below=of smallk] (largek) {Large $k$\\Practicality};
\node[box=OI3, below=of largek] (fairness) {Fairness\\Relaxation Pitfall};
\node[box=OI3, below=of fairness] (nonneg) {Non-negativity\\Trap};
\node[inner sep=0pt, above=of smallk] (alg) {\Large\textbf{Algorithm Design}};
\begin{scope}[on background layer]
\node[cluster=OI3, fit=(alg)(largek)(nonneg)(fairness)(smallk), anchor=center] (alg_cluster) {};
\end{scope}

\node[box=OI1, below right= 1pt and 110pt of largek] (khd) {Multi-Dim.\\$k$-level-based};
\node[box=OI1, above= of khd] (k2d) {2-D $k$-level-based};
\node[box=OI1, below= 16pt of khd] (milp) {MILP-based};
\node[inner sep=0pt, above=20pt of  k2d] (pract) {\Large\textbf{Practical Algorithms}};
\node[subcluster=OI1, fit=(khd)(k2d)] {};
\begin{scope}[on background layer]
\node[cluster=OI1, fit=(pract)(milp)(khd)(k2d)] (pract_cluster){};
\end{scope}

\node[box=OI4, above right= 30pt and 60pt of k2d] (tie) {Tie-breaking};
\node[box=OI4, above = of tie] (hardware) {Hardware Utilization};
\node[box=OI4, above = of hardware] (space) {Space Overhead};
\node[inner sep=0pt, above=of space, text width=80pt, align=center] (eng) {\Large \textbf{Engineering\\Challenges}};
\begin{scope}[on background layer]
\node[cluster=OI4, fit=(eng)(tie)(hardware)(space)] (eng_cluster){};
\end{scope}

\node[box=OI5, right=200pt of khd] (runhd) {Multi-Dim.\\Runtimes};
\node[box=OI5, above= of runhd] (run2d) {2-D Runtimes};
\node[box=OI5, below=16pt of runhd] (quality) {Quality\\Evaluations};
\node[inner sep=0pt, above=20pt of  run2d] (exp) {\Large\textbf{Experiments}};
\node[subcluster=OI5, fit=(run2d)(runhd)] {};
\begin{scope}[on background layer]
\node[cluster=OI5, minimum width = 120pt, fit=(exp)(quality)(runhd)(run2d)] (exp_cluster) {};
\end{scope}

\draw [very thick, -{Latex[length=3mm]}] ([yshift=-20pt]hardness_cluster.east) -- ([yshift=-4pt]alg_cluster.west) node[midway,above] {\Large guides};
\draw [very thick, -{Latex[length=3mm]}] ([yshift=-2pt]alg_cluster.east) -- ([yshift=6pt]pract_cluster.west) node[midway,above] {\Large drives};
\draw [very thick, -{Latex[length=3mm]}] (pract_cluster.north) -- (eng_cluster) node[midway,above, sloped] {\Large address};
\draw [very thick, -{Latex[length=3mm]}] ([yshift=8pt]pract_cluster.east) -- ([yshift=8pt]exp_cluster.west) node[midway,above] {\Large undergo};
\draw [very thick, -{Latex[length=3mm]}, dashed] (exp_cluster.north) -- (eng_cluster) node[midway,above,sloped] {\Large expose};
\draw [very thick, -{Latex[length=3mm]}, dashed] (exp_cluster.south) |-|[ratio=5.0] (alg_cluster.south) node[midway, above] {\Large refine};
\end{tikzpicture}
}
    \caption{The structure and interplay of key results and components in this work. Solid arrows denote the primary workflow and direct influences, while dashed arrows indicate feedback. Dashed boxes group related components presented in the same subsection.}\label{fig:structure}
\end{figure}

\subparagraph{Reader's guide.} As mentioned, our solution stems from a tight integration of theoretical
and practical work, and readers would benefit from considering both perspectives while reading the paper. However, we also offer a guide for readers with specific interests. Practically-oriented readers may skip the hardness proofs in Section~\ref{sec:hardness} and focus on the implications presented in the Remarks~\ref{rmk:3sum}--\ref{rmk:np} within that section, using Section~\ref{subsec:implications} as a guide. Theoretically-oriented readers may find the engineering challenges outlined in Section~\ref{subsec:practical_klevel} informative, but may skip the corresponding implementation details and use Section~\ref{subsec:choice} to navigate the experimental results in Sections~\ref{subsec:expruntime} and~\ref{subsec:expquality}.

\section{Preliminaries}
\label{sec:prel}
In the Fair Top-$k$ Selection problem, we are given a set, $\mathcal{C}$, of candidates (or items), each described by $d$ scoring attributes and a sensitive  attribute. For each $c\in \mathcal{C}$, we use a point $p(c) = (p_1, p_2, \dots, p_d)$ in $\mathbb{R}^d$ to represent the values of its scoring attributes and a label $A(c)$ to represent the value of its sensitive attribute, where $A(c) \in \{\mathcal{G}_1,\mathcal{G}_2,\dots,\mathcal{G}_m\}$. Among these labels, $\mathcal{G}_1$ denotes the protected group, which indicates a membership in a minority or historically disadvantaged group. With a slight abuse of notation (to align with \cite{zehlike2022fairness}), we also use $\mathcal{G}_i$ to denote the subset of candidates in $C$ whose sensitive attribute value is the label $\mathcal{G}_i$, particularly for case of $i = 1$. To simplify notation, we may use $p$ instead of $p(c)$, if there is no ambiguity. Besides, we may refer to a candidate $c$ by its corresponding point $p(c)$ when the context is clear.

We focus on the class of linear scoring functions. Such a function is given by a real-valued weight vector $w = (w_1, w_2, \dots, w_d)$, where each $w_i\geq 0$. (This non-negativity of linear scoring functions is a widely adopted assumption, e.g., in \cite{vlachou2010reverse, cao2017k, asudeh2019designing}.) The score of a candidate $c$ is computed as $f(p(c)) = w\cdot p(c) = \sum_{i=1}^d w_ip_i$. The top-$k$ candidates for a given $w$ are the candidates with the $k$ highest scores. In the Fair Top-$k$ Selection problem, the goal is to find a weight vector for which the top-$k$ candidates satisfy the fairness constraint (described below).

At this point it is worth noting that for a given $w$ and $k$, the top-$k$ subset may not be unique due to ties in scores. In this case, we consider all top-$k$ subsets and regard $w$ as a fair weight vector if any of the top-$k$ subsets satisfies the fairness constraint. This is embodied in the following definition:
\begin{mydef}
    For a given weight vector $w$ and a positive integer $k \leq n$, a subset $\tau_k^w \subseteq C$ is a top-$k$ subset of $C$ if $|\tau_k^w| = k$ and for any pair of candidates that $c \in \tau_k^w$ and $c' \in C \setminus \tau_k^w$, $w \cdot p(c) \geq w \cdot p(c')$.
    \label{def:topk}
\end{mydef}

We now formalize our notion of fairness. Let $\tau_k^w$ denotes a top-$k$ subset for $w$. Let $L_{k}^{\scriptscriptstyle \mathcal{G}_1}$ (resp., $U_{k}^{\scriptscriptstyle \mathcal{G}_1}$) denotes a given lower (resp., upper) bound on the number of candidates of the protected group desired in the top-$k$ subset. Ideally, these should be close to $k \cdot (|\mathcal{G}_1|/n)$, i.e., the candidates of $\mathcal{G}_1$ are represented in roughly the same proportion in the top-$k$ subset as they are in the entire dataset. In other words, $w$ is a fair scoring weight vector if 
\begin{equation}
    L_{k}^{\scriptscriptstyle \mathcal{G}_1} \leq | \tau_k^w \cap \mathcal{G}_1 | \leq U_{k}^{\scriptscriptstyle \mathcal{G}_1}.
\end{equation}

Finally, one may wish to limit the search for a fair weight vector to a subset of the weight vector space. We denote this subset by $V$ and describe it via a set of $l$ linear inequalities. Furthermore, we assume $||w||_1 = \sum_{i=1}^d w_i = 1$ (recall each $w_i \geq 0$), as normalizing a non-negative weight vector by its $L_1$ norm still preserves the relative order between candidates. Note that with $||w||_1 = 1$, $V$ can be regarded of $\mathbb{R}^{d-1}$ as $w_d$ can be uniquely determined by $w_d = 1 - \sum_{i=1}^{d-1}w_i$. Our aim is to find a weight vector $w \in V$ which selects a top-$k$ subset that is fair. 

Formally, we define our problem as follows:  

\begin{pro}[Fair Top-$k$ Selection]
    We are given a set of $n$ candidates with $d$ scoring attributes and one sensitive attribute. Each candidate, $c$, is represented by a point $p$ in $\mathbb{R}^d$ and a label $A(c) \in \{\mathcal{G}_1,\mathcal{G}_2,\dots,\mathcal{G}_m\}$. We are also given a non-negative integer $k \leq n$, non-negative integers $L_{k}^{\scriptscriptstyle \mathcal{G}_1}$ and $U_{k}^{\scriptscriptstyle \mathcal{G}_1}$, and a subset $V$ of the weight vector space described by $l$ linear inequalities. The goal is to find a real-valued weight vector $w =(w_1, w_2, \ldots, w_d) \in V$, where $w_i\geq 0$ for all $i$ and $||w||_1 = 1$, such that
    \begin{equation*}
        L_{k}^{\scriptscriptstyle \mathcal{G}_1} \leq | \tau_k^w \cap \mathcal{G}_1 | \leq U_{k}^{\scriptscriptstyle \mathcal{G}_1}.
    \end{equation*}
\end{pro}

Note that for the flexibility, our problem definition does not require an initial unfair weight vector, as in the case of Example~\ref{eg:admission}. Besides, even though our formulation allows $p(c)$ to be any point $\mathbb{R}^d$, in practice, values of all scoring attributes are typically normalized to $[0,1]^d$. This is justified for two reasons: (1) Uniformly translating the entire point set (induced by scoring attributes) preserves the relative order between two candidates for any weight vector, allowing all values of scoring attributes to be made non-negative. (2) Scaling one score attribute, e.g., transforming a point $p=(p_1, p_2, \ldots, p_i, \ldots, p_d)$ to $(p_1, p_2, \ldots, \eta p_i, \ldots, p_d)$, can be canceled out by scaling the corresponding weight of the weight vector from $w=(w_1,w_2, \ldots, w_i, \ldots, w_d)$ to $(w_1, w_2, \ldots, (1/\eta)w_i, \ldots, w_d)$. Therefore, normalization is often used in practice to make weights comparable across attributes.

\section{Hardness}\label{sec:hardness}
Known algorithms for the Fair Top-$k$ Selection problem \cite{asudeh2019designing} do not scale well with the dataset size, especially in higher dimensions. Here we establish (conditional) lower bounds suggesting that the problem is unlikely to be efficiently solvable for multi-dimensional data, especially as the dimensionality increases. We first consider a special case of the problem in fixed dimensions and then extend our results to any reasonable value of  $L_{k}^{\scriptscriptstyle \mathcal{G}_1}$ and $U_{k}^{\scriptscriptstyle \mathcal{G}_1}$ as well as to any $k = \beta n$, where $0 < \beta < 1$ is a constant. Finally, we show the problem is NP-hard in arbitrary dimensions. In particular, the practical implications of these hardness results are discussed in Remarks~\ref{rmk:3sum}--\ref{rmk:np}.

\subsection{Hardness results in fixed dimensions}\label{sec:hardness_fixed}
We show that for any fixed (i.e., constant) dimension, $d\geq 3$, the Fair Top-$k$ Selection problem is 3SUM-Hard \cite{williams2018some, gajentaan1995class}, hence is unlikely to be solvable in $O(n^{2-\delta})$ time for any constant $\delta > 0$. 3SUM is a well-known computational problem and its hardness assumption plays a key role in  fine-grained complexity theory \cite{williams2018some}. Furthermore,  based on a result in  \cite{erickson1999new}, we show that the problem is unlikely to be solvable in $O(n^{d-1})$ time for any $d\geq3$. All our conditional lower bounds are established for a special case of the problem, where there is only one candidate of the protected group and there are no constraints on the weight vector. The problem then becomes finding a weight vector $w$ such that the only candidate of the protected group is in the top-$k$ subset.

Denote the only candidate of the protected group by $t$ and its values of scoring attributes $p(t)$ by $q$. By  definition, $t$ is in the top-$k$ subset for a weight vector $w$ if and only if there are at least $n-k$ candidates $c\in C \setminus \{ t \}$ such that $w \cdot p(c) \leq w \cdot p(t)$. The problem now becomes that of finding a weight vector $w$ satisfying at least $n-k$ of the $n-1$ inequalities $w \cdot p(c) \leq w \cdot q$. 

Deciding whether three points in a given set in $\mathbb{R}^2$ lie on a line (known as the point collinearity problem) is known to be 3SUM-Hard~\cite{gajentaan1995class}. Here we give a reduction from the collinearity problem to our problem, thereby showing that it, too, is 3SUM-Hard.

\begin{thm}[3SUM-hardness]\label{thm:3sum}
    For any $d\geq 3$, Fair Top-$k$ Selection cannot be solved in $O(n^{2-\delta})$ time for any constant $\delta > 0$, assuming the 3-SUM hypothesis.
\end{thm}

\begin{proof}
    The proof is given only for  $d = 3$ here, as generalizing it to higher dimensions is straightforward.

    Let $P = \{(x_1, y_1), (x_2, y_2),\cdots, (x_n, y_n) \}$ be an arbitrary instance of the  collinearity problem. A line in $\mathbb{R}^2$ is described as a function $ax + by + c = 0$. Deciding collinearity is equivalent to finding a triple of real numbers, $(a, b, c)$, such that at least three of the $n$ linear equations below are satisfied:
    \begin{equation}
        \left\{\begin{aligned}
            ax_1 + by_1 + c &= 0,\\
            ax_2 + by_2 + c &= 0,\\
            & \vdotswithin{ = } \\
            ax_n + by_n + c &= 0.
          \end{aligned}\right.
          \label{eq:n}
    \end{equation}

    For $1 \leq i \leq n$, we replace the $i$th equation above by two inequalities: $ax_i + by_i + c \leq 0$ and $-ax_i - by_i - c \leq 0$. This yields a system of $2n$ linear inequalities:
    \begin{equation}
        \left\{\begin{aligned}
            ax_i + by_i + c & \leq 0,\, 1 \leq i \leq n,\\
            -ax_i - by_i - c &\leq 0,\, 1 \leq i \leq n.\\
          \end{aligned}\right.
          \label{ineq:2n}
    \end{equation}
    Consider a triple $(a, b, c)$. For each $i$, we have either  $ax_i + by_i + c = 0$ or $ax_i + by_i + c \not= 0$ in~(\ref{eq:n}). If the former, then  $ax_i + by_i + c \leq 0$ and $-ax_i - by_i - c \leq 0$. If the latter, then $ax_i + by_i + c < 0$ and $-ax_i - by_i - c > 0$, or vice versa. In any case, for each $i$, at least one of the two inequalities in~(\ref{ineq:2n}) must hold. It follows that, if three or more of the $n$ points in $P$ are collinear, then there is a triple $(a, b, c)$ such that at least $2\times 3 + (n-3)=n+3$ of the inequalities in~(\ref{ineq:2n}) are satisfied. If no three points in $P$ are collinear, then at most $2\times 2 + (n-2)=n+2$ of the inequalities in~(\ref{ineq:2n}) are satisfied. 

    Next, we create an instance of the Fair Top-$k$ Selection problem, as follows. Let $q=(0,0,0)$ be the origin of the coordinate system. Let $P' = \{(x_i, y_i, 1), (-x_i, -y_i, -1) \mid 1 \leq i \leq n\}\cup\{q\}$ be the set of $2n+1$ points in $\mathbb{R}^3$ obtained from $q$ and $P$. Point $q$ is assigned to the protected group $\mathcal{G}_1$ and all other points in $P'$ are assigned arbitrarily to other groups. Finally, we set $k=n-2$ and $L_{k}^{\scriptscriptstyle \mathcal{G}_1}  = U_{k}^{\scriptscriptstyle \mathcal{G}_1} = 1$. This is the instance of the Fair Top-$k$ Selection problem that we seek. Clearly, the reduction takes $O(n)$ time.
    
    Suppose that $q$ is in the top-$k$ subset for some weight vector $w$. Then there are $2n+1-k=n+3$ points $p\in P'\setminus \{q\}$ that are not in the top-$k$ subset. By Definition~\ref{def:topk}, $p\cdot w \leq q \cdot w = 0$. Additionally, there could be points $p$ in the top-$k$ subset such that $p\cdot w \leq 0$. Thus, there are at least $n+3$ points $p$ for which $p\cdot w \leq 0$. Let $(a,b,c)$ be the triple corresponding to the components of $w$ (with $||w||_1=a+b+c=1$ ensuring $w$ is not a zero vector) and note that each $p\in P'\setminus \{q\}$ is of the form $(x_i, y_i, 1)$ or $(-x_i, -y_i, -1)$ for $1\leq i \leq n$. It follows that at least $n+3$ inequalities in~(\ref{ineq:2n}) are satisfied, so there are at least three points of $P$ on a line. 
    
    On the other hand, suppose that $q$ is not in the top-$k$ subset for any $w$. Then any top-$k$ subset contains $k$ points $p\in P'\setminus \{q\}$ with $p\cdot w > q \cdot w = 0$. There are $2n+1-k-1=n+2$ points $p\in P'\setminus \{q\}$ that are not in the top-$k$ subset, hence at most $n+2$ points $p\in P'\setminus \{q\}$ have $p\cdot w \leq 0$. Thus, at most $n+2$ of the inequalities in~(\ref{ineq:2n}) are satisfied, so no three points of $P$ are on a line. It follows that point collinearity for $P$ can be decided from a solution to the Fair Top-$k$ Selection problem.  

    There is one subtlety that we need to discuss.
    Recall that the Fair Top-$k$ Selection problem requires the components $a, b, c$ of $w$ to be non-negative. If there are three points of $P$ on a line $ax+by+c=0$ where (say) $a<0$ and $b, c \geq 0$, then this triple will be missed when the Fair Top-$k$ Selection algorithm is run on the derived dataset $P'$ in $\mathbb{R}^3$. Fortunately, this case can be handled as follows: We create a new point set $\widehat{P} = \{(-x_1, y_1), (-x_2, y_2),\cdots, (-x_n, y_n) \}$ and use this to derive the dataset in $\mathbb{R}^3$ on which Fair Top-$k$ Selection can be run. The collinear points will then be found  since the first coordinate of each point in $\widehat{P}$ is the negative of its value in $P$. Similarly, we create two more datasets corresponding to $a, c\geq 0, b <0$ and $a,b < 0, c\geq 0$. Thus, besides $P$, we have three additional datasets and we run the Fair Top-$k$ Selection algorithm on all four. (The case $c<0$ need not be considered since a line $ax+by+c=0$ with $c<0$ is equivalent to $-ax-by-c=0$ and lines of the latter type have been handled by the newly created sets.)  
\end{proof}

\begin{rmk}\label{rmk:3sum}
    For large datasets, practical efficiency typically requires algorithms with run times $O(n\cdot \textnormal{\text{polylog}}(n))$. However, since the 3SUM problem is widely conjectured to have a lower bound of $\Omega(n^{2-\delta})$ (for any constant $\delta > 0$) in the Word-RAM mode \cite{williams2018some}, our reduction suggests that such an efficient algorithm is unlikely to exist for general problem instances in dimensions $d\geq 3$.
\end{rmk}

Deciding whether three points are on a line is a special case of the Affine Degeneracy problem in $\mathbb{R}^d$, namely deciding whether  $d+1$ points from a given point set in $\mathbb{R}^d$ are on a  hyperplane. Using a similar approach as above the following result can be shown.

\begin{col}[Curse of Dimensionality]\label{col:hd}
    For any constant $d \geq 3$, Fair Top-$k$ Selection in $\mathbb{R}^d$ is at least as hard as the Affine Degeneracy problem in $\mathbb{R}^{d-1}$.
\end{col}

\begin{proof}
    For a given instance of the Affine Degeneracy problem in $\mathbb{R}^{d-1}$, we construct $2^{d-1}$ instances of the Fair Top-$k$ Selection problem in $\mathbb{R}^d$. Each instance is created using the method in the proof of Theorem \ref{thm:3sum}, with the resulting input candidate set of size $2n + 1$. Then, to decide whether there exists $d$ points lying on a hyperplane, one solves each instance of the Fair Top-$k$ Selection problem with $k = n - d + 1$ and $L_{k}^{\scriptscriptstyle \mathcal{G}_1}  = U_{k}^{\scriptscriptstyle \mathcal{G}_1} = 1$. The reduction takes $O(n)$ time with $d$ being a constant.
\end{proof}

\begin{rmk}\label{rmk:hd}
    In a certain decision tree model, all algorithms for the Affine Degeneracy problem require $\Omega(n^d)$ time~\cite{erickson1999new}. This suggests a lower bound of $\Omega(n^{d-1})$ for Fair Top-$k$ Selection in $\mathbb{R}^d$ for $d \geq 3$. The result shows that as $d$ increases, the lower bound of the problem grows exponentially, indicating that the computational complexity increases significantly in higher dimensions. However, the reduction only applies for $d$ being a constant, since the number of constructed instances also grows exponentially with $d$ (see Section \ref{hardness:ad} for further discussion).
\end{rmk}

\subsection{Hardness results for more general cases}\label{sec:hardness_general}
As was stated, the conditional lower bound in fixed dimensions is established using a special case of the problem, where $k$ is set to be $ (n - 1)/2 - d + 1$ (notice that in the proof of Corollary \ref{col:hd}, we used an instance of size $2n + 1$ and $k = n - d + 1$ for the reduction), $C$ has only one candidate of $\mathcal{G}_1$, and the fairness constraint must be $L_{k}^{\scriptscriptstyle \mathcal{G}_1}  = U_{k}^{\scriptscriptstyle \mathcal{G}_1} = 1$. However, for a typical instance of the Fair Top-$k$ Selection problem in practice, the value of $k$ is unlikely to be close to half the dataset size since one usually wants to select a small subset. Also, it is unlikely that there is only one candidate of $\mathcal{G}_1$. Additionally, it is unlikely to have $L_{k}^{\scriptscriptstyle \mathcal{G}_1}$ and $U_{k}^{\scriptscriptstyle \mathcal{G}_1}$ be exactly the same, as intuitively, finding a weight vector that satisfies fairness constraint with some ``relaxation'' is easier. Finally, for a fair top-$k$ selection, $L_{k}^{\scriptscriptstyle \mathcal{G}_1}$ and $U_{k}^{\scriptscriptstyle \mathcal{G}_1}$ are also expected to be close to $k \cdot (|\mathcal{G}_1|/n)$. Here, we show that our conditional lower bounds also hold when the value of $k$, the number of points of $\mathcal{G}_1$ and the fairness constraint are more general and realistic.

We begin by introducing several key concepts. Given a point set $S \subseteq \mathbb{R}^d$, a point $q \in \mathbb{R}^d$ is a \textit{dominating} (resp. \textit{dominated}) point of $S$ if for any point $p \in S$, $q_i > p_i$ (resp. $q_i < p_i$) for all $1 \leq i \leq d$. One can verify that for any non-negative weight vector, the score of a dominating (resp. dominated) point will be larger (resp. smaller) than the score of any point in $S$. Also, it is easy to generate dominating or dominated points for a given $S$. One only has to look for the highest (resp. lowest) value for each coordinate in the points of $S$. We say a candidate $c$ is a \textit{dominating} (resp. \textit{dominated}) candidate of a candidate set $C$ if $p(c)$ is a dominating (resp. dominated) point of the point set induced by the scoring attributes of $C$.

With these concepts, we first investigate how to extend our hardness results to a general $k$ value. Suppose that we have an efficient algorithm that can only handle $k = \beta n$, where $0 < \beta < 1$ is a fixed constant. We show that it can also be applied to solve the $k  = (n - 1)/2 - d + 1$ instance efficiently.

\begin{thm}[$k$-Generalization]\label{thm:k}
    Given a polynomial-time algorithm for Fair Top-$k$ Selection with $k = \beta n$, where $0 < \beta < 1$ is a fixed constant, one can solve the Fair Top-$k$ Selection with $k  = (n - 1)/2 - d + 1$ in the same asymptotic run time.
\end{thm}

\begin{proof}
    Since $k$ is expected to be small in practice, we consider the case where $k$ is small first, that is, $\beta n  < (n - 1)/2 - d + 1$. For the given candidate set $C$, we construct $C'$ from $C$ by adding a number of dominated candidates, making the size of $C'$ to be $(n - 1)/2\beta - (d-1)/\beta$. $A(c)$ of a candidate $ c\in  C' \setminus C$ can be arbitrarily assigned. Then, we run the algorithm for Fair Top-$k$ Selection with $k = \beta n$ on $C'$. Given the size of $C'$, we have $k  = (n - 1)/2 - d + 1$. Since all candidates in $C' \setminus C$ are dominated ones, none of these candidates will be in the top-$k$ subset for any $k \leq n$. So, the top-$k$ selection result will only contain candidates in $C$.Since $\beta$ is a constant, the size of $C'$ is still $O(n)$ and the additional time spent on constructing $C'$ is will not change the asymptotic run time. 
    
    The case $\beta n > (n - 1)/2 - d + 1$ can be similarly handled. Let $\gamma = \beta n - ((n - 1)/2 - d + 1)$, we construct $C'$ from $C$ by adding $n'$ dominating candidates where $n'= \gamma / (1 - \beta)$. $A(c)$ of a candidate $c \in C' \setminus C$ is assigned arbitrarily to any group other than $\mathcal{G}_1$. For $C'$, any top-$\beta n$ subset must contain all candidates in  $c \in C' \setminus C$, given $\beta|C'| = \beta (n + n') = (n - 1)/2 - d + 1 + n'$. By removing all candidates in $C' \setminus C$ in the top-$k$ subset, one can recover any top-$k$ subset (where $k = (n - 1)/2 - d + 1$) of $C$ from the corresponding top-$\beta n$ subset of $C'$. Since all candidates in $C' \setminus C$ are not of $\mathcal{G}_1$, the fair weight vector found by running the given algorithm on $C'$ will also be a fair one for $C$, with the fairness constraint remaining the same. Conversely, if there does not exist a fair weight vector on $C$, the given algorithm will also report a failure.
\end{proof}

\begin{rmk}\label{rmk:k}
    Since more efficient algorithms for cases where $k = \beta n$ would imply a violation of lower bounds established in Section \ref{sec:hardness_fixed}, this reduction extends those lower bounds to a broad range of $k$. However, notice that for a sufficiently small $k$ (e.g., $k = O(\textnormal{\text{polylog}}(n))$), the reduction breaks down as $C'$ requires adding too many dominated candidates, causing the reduction to take more than $O(n)$ time. This reveals that the previous lower bounds may not be applicable when $k$ is sufficiently small, a critical insight that informs our algorithm design.
\end{rmk}

Next, we extend the result to a larger number of points of the protected group and to more general fairness constraint. The proof also starts by considering a simple case.  Consider a dataset in which all candidates belong to one of the two group and where the number of candidates of the protected group, $n_1$, and the number of candidates that of the other group, $n_2$, are equal. Suppose we have an efficient Fair Top-$k$  Selection algorithm which can only handle datasets where $n_1 = n_2 = n/2$, and the fairness constraint must be set to $L_{k}^{\scriptscriptstyle \mathcal{G}_1}  = U_{k}^{\scriptscriptstyle \mathcal{G}_1} = k \cdot (n_1/n) = k/2$, that is, the proportion of candidates of $\mathcal{G}_1$ in the top-$k$ subset is the same as the proportion in the entire dataset. We show that with such an algorithm, one can solve the case where $n_1$ = 1 and $L_{k}^{\scriptscriptstyle \mathcal{G}_1}  = U_{k}^{\scriptscriptstyle \mathcal{G}_1} = 1$ in the same asymptotic run time.

\begin{thm}[Group-balancing]\label{thm:perfect}
    Given a polynomial-time algorithm for Fair Top-$k$ Selection with $n_1 = n_2 = n/2$ and $L_{k}^{\scriptscriptstyle \mathcal{G}_1}  = U_{k}^{\scriptscriptstyle \mathcal{G}_1} = k/2$, one can solve the Fair Top-$k$ Selection with $n_1 = 1$ and $L_{k}^{\scriptscriptstyle \mathcal{G}_1}  = U_{k}^{\scriptscriptstyle \mathcal{G}_1} = 1$ in the same asymptotic run time.
\end{thm}

\begin{proof}
    Since the given algorithm for the reduction only accepts a dataset where $n_1 = n_2$, to solve the case where $n_1 = 1$ and $n_2 = n - 1$, the first step is to add more points into the dataset so that the given algorithm can accept it (assuming $n$ is sufficiently large). Similar to what was done in the proof of Theorem \ref{thm:k}, We construct a new dataset $C'$ from  $C$ by adding $n - 2$ of dominated candidates of $\mathcal{G}_1$. Now we have a dataset $C'$ where $n_1 = n_2 = n - 1$.
    
    The given algorithm accepts $C'$, however, it also forces $L_{k}^{\scriptscriptstyle \mathcal{G}_1}$ and $U_{k}^{\scriptscriptstyle \mathcal{G}_1}$ to be $k/2$. Since all candidates in $C' \setminus C$ are dominated, the given algorithm will always fail to produce a fair weight vector for any $4 \leq k < n$ as $C$ only has $1$ candidate of $\mathcal{G}_1$. To address this issue, we construct another candidate set $\widehat{C}$ from $C'$ by substituting some dominated candidates in $C' \setminus C$ with dominating candidates. For any input $k = a$ where $1 \leq a < n$ ($k = n$ is trivial since a candidate will always be among the top-$n$), we construct $\widehat{C}$ from $C'$ by arbitrarily choosing $a - 1$ dominated candidates in $C' \setminus C$ and substituting them with dominating ones (with the group membership unchanged). To find a weight vector where the only candidate of $\mathcal{G}_1$ in $C$ is among top-$k$, one only has to run the given Fair Top-$k$ Selection algorithm on $\widehat{C}$ with $k = 2a$ and $L_{k}^{\scriptscriptstyle \mathcal{G}_1} = U_{k}^{\scriptscriptstyle \mathcal{G}_1} = a$. Since there are $a - 1$ candidates of $\mathcal{G}_1$ in $\widehat{C}$ that dominate all other candidates, any top-$2a$ subset must contain all of them. If the algorithm successfully finds a weight vector, the one additional candidate of $\mathcal{G}_1$ in the top-$2a$ subset must be the one in $C$ since all other candidates of $\mathcal{G}_1$ in $\widehat{C}$ expect the $a - 1$ dominating ones are dominated. On the contrary, if there does not exist a top-$a$ subset containing the only candidate of $\mathcal{G}_1$ in $C$, the given algorithm will also report a failure.
    
    The additional time spent on constructing $\widehat{C}$ is $O(dn)$, same as the time spent on reading inputs. Also, since the run time of the given algorithm is polynomial, doubling the input $k$ value (from $a$ to $2a$) will not increase the asymptotic run time.
\end{proof}

Theorem \ref{thm:perfect} still has a restriction on fairness constraint that $L_{k}^{\scriptscriptstyle \mathcal{G}_1}$ and $U_{k}^{\scriptscriptstyle \mathcal{G}_1}$ must be exactly equal. In practice, one typically sets $L_{k}^{\scriptscriptstyle \mathcal{G}_1} =  k\cdot (n_1/n) - \sigma_L$  and $U_{k}^{\scriptscriptstyle \scriptscriptstyle \mathcal{G}_1} = k\cdot (n_1/n) + \sigma_U$, where $\sigma_L$ and $\sigma_U$ are non-negative integers. By carefully manipulating the number of dominated and dominating candidates, we are able to show the following:

\begin{col}[Constraint Relaxation]\label{col:relaxed}
    Given a poly-nomial-time algorithm for Fair Top-$k$ Selection with $L_{k}^{\scriptscriptstyle \mathcal{G}_1} = k\cdot (n_1/n) - \sigma_L$ and $U_{k}^{\scriptscriptstyle \mathcal{G}_1} = k\cdot (n_1/n) + \sigma_U$, where $\sigma_L$ and $\sigma_U$ are non-negative integers,  one can solve the Fair Top-$k$ Selection with $n_1 = 1$ and $L_{k}^{\scriptscriptstyle \mathcal{G}_1}  = U_{k}^{\scriptscriptstyle \mathcal{G}_1} = 1$ in the same asymptotic run time.
\end{col}

\begin{proof}
    We first consider the case where $n_1 = n_2$. If $L_{k}^{\scriptscriptstyle \mathcal{G}_1}$ can be set to $1$, one only has to construct $C'$ from $C$ by adding $n - 2$ dominated candidates of the protected group $\mathcal{G}_1$ and run the given algorithm on $C'$. Since all candidates in $C' \setminus C$ are dominated, top-$k$ subsets of $C'$ and $C$ are equivalent for any $1 \leq k < n$. A top-$k$ subset (with $1 \leq k < n$) of $C'$ satisfies the fairness constraint if and only if it contains the only candidate of $\mathcal{G}_1$ in $C$. For the case where $L_{k}^{\scriptscriptstyle \mathcal{G}_1}$ must be greater than $1$, similar to what we did in the proof of Theorem \ref{thm:perfect}, for an input $k = a$ where $1 \leq a < n$, we construct $\widehat{C}$ by arbitrarily choosing $a - 1 - \sigma_L$ candidates in $C' \setminus C$ and substituting them with dominating ones. Then, we run the given algorithm on $\widehat{C}$ with $k = 2a$, $L_{k}^{\scriptscriptstyle \mathcal{G}_1} = a - \sigma_L$  and $U_{k}^{\scriptscriptstyle \mathcal{G}_1} = a + \sigma_U$. Since those $a - 1 - \sigma_L$ dominating candidates of $\mathcal{G}_1$ must be in all top-$k$ subsets for any $1 \leq k < n$, finding a weight vector whose corresponding top-$k$ subset of $\widehat{C}$ satisfies the fairness constraint is equivalent to finding a weight vector whose corresponding top-$k$ subset contains the only candidate of $\mathcal{G}_1$ in $C$. Note that the value $U_{k}^{\scriptscriptstyle \mathcal{G}_1}$ is in fact irrelevant in this reduction since any top-$k$ satisfying the fairness constraint contain at most $L_{k}^{\scriptscriptstyle \mathcal{G}_1}$ candidates of $\mathcal{G}_1$.

    The case where $n_1 \neq n_2$ is also similar. If $L_{k}^{\scriptscriptstyle \mathcal{G}_1}$ can be set to $1$, one only has to construct $C'$ by adding $n'$ dominated candidates of the $\mathcal{G}_1$ such that $(n' + 1)/(|C| - 1) = n_1/n_2$ and run the given algorithm on $C'$. For the case where  $L_{k}^{\scriptscriptstyle \mathcal{G}_1} > 1$, one constructs $\widehat{C}$ from $C'$ by substituting  $a - 1 - \sigma_L$ dominated candidates in $C' \setminus C$ with dominating ones so that finding a fair weight vector on $\widehat{C}$ with $k = a \cdot (n/n_1)$ (where $n = |\widehat{C}|$ and $n_1 = |\widehat{C} \cap \mathcal{G}_1|$), $L_{k}^{\scriptscriptstyle \mathcal{G}_1} = a - \sigma_L$  and $U_{k}^{\scriptscriptstyle \mathcal{G}_1} = a + \sigma_U$ is equivalent to finding a weight vector whose corresponding top-$k$ subset contains the only candidate of $\mathcal{G}_1$ in $C$.
\end{proof}

\begin{rmk}\label{rmk:relaxed}
    Intuitively, relaxing the fairness constraint should make the problem computationally easier, as it becomes less restrictive to find a fair weight vector. However, our reduction indicates that more efficient algorithms for solving the problem under the relaxed fairness constraint would imply a violation of the lower bounds established in Section~\ref{sec:hardness_fixed}, which is unlikely.
\end{rmk}

\subsection{Hardness result in arbitrary dimensions}\label{hardness:ad}
Our $\Omega(n^{d-1})$ conditional lower bound is based on a certain decision tree model, which is a restricted computational model. To strengthen our claim, we show that the problem is NP-hard in arbitrary dimensions, which also explains why the problem becomes harder to solve as the dimensionality increases.

With Corollary \ref{col:hd}, one might hope to use the Affine Degeneracy problem and our technique in the fixed dimensions to show the problem is NP-hard in arbitrary dimensions, since the Affine Degeneracy problem is known to be NP-hard in arbitrary dimensions \cite{erickson1999new}. However, a straightforward extension will not work since using this technique, one has to solve $2^{d-1}$ instances of the Fair Top-$k$ Selection problem in $\mathbb{R}^d$ to solve the Affine Degeneracy problem in $\mathbb{R}^{d-1}$. In arbitrary dimensions, this is not a polynomial time reduction. To resolve this issue, we use an alternative problem, Maximum Independent Set, to establish the NP-hardness result, starting from its binary integer programming formulation \cite{lovasz1979shannon}. Again, we use the special case of our problem, that there is only one candidate of the protected group and there are no constraints on the weight vector, for the reduction. (It can be easily extended to more general cases by results in Section \ref{sec:hardness_general}.)

\begin{thm}[NP-hardness]\label{thm:np}
    Fair Top-$k$ Selection is NP-hard in arbitrary dimensions.
\end{thm}

\begin{proof}
    For the reduction, we transform the problem into finding a solution satisfying maximum number of linear inequalities. To obtain the system of linear inequalities, we start by formulating this problem into a binary integer program as it was done in \cite{lovasz1979shannon}.

    Consider a graph with $n$ nodes and $m$ edges, where $E$ denotes the set of edges. Let $x_i\in \{0, 1\}$ be a binary variable where $x_i = 1$ if the node $i$ is in the independent set and $x_i = 0$ otherwise. One can solve the Max Independent Set problem by solving a binary integer program that maximizes $\sum_{i = 1}^n x_i$ and is subject to constraints $x_i + x_j \leq 1$ for each edge $\{i, j\} \in E$. 

    With the binary integer program as a starting point, each $x_i$ is relaxed with the introduction of a real variable $x_{n+1} \in [0,1]$ so that it can also take a real value in $[0,1]$. Now, we have a new set of constraints that $x_i + x_j - x_{n+1} \leq 0$ for each $\{i, j\} \in E$. For each of these inequalities, we make $n$ variants of them which result in a set of constraints that $bx_i + bx_j - bx_{n+1} \leq 0$ for each integer $1 \leq b \leq n$ and each $\{i, j\} \in E$. Finally, we remove the objective function and add a new set of constraints that  $x_{n+1} \geq x_i$ for each node $i$. Now, we have the following system of $nm + n$ linear inequalities:
    \begin{equation}\label{eq:nphard}
        \left\{\begin{aligned}
            bx_i + bx_j - bx_{n+1}& \leq 0, 1 \leq b \leq n \text{ and } \{i, j\} \in E,\\
            -x_i + x_{n+1}&\leq 0,\, 1 \leq i \leq n.
          \end{aligned}\right.
    \end{equation}

  Similar to what was done in the proof of Theorem \ref{thm:3sum}, we then transform the system of inequalities into a candidate set of size $nm + n + 1$, with one candidate $t$ (the origin of the coordinate system for $\mathbb{R}^{n+1}$) assigned to the protected group $\mathcal{G}_1$ and others arbitrarily assigned to other groups. We run the Fair Top-$k$ Selection algorithm for all $1 \leq k \leq n$ and keep the solution with the smallest $k$ value. For the (non-negative and real) resulting vector $x = (x_1, x_2, \dots, x_{n+1})$ (with $||x||_1 = 1$), we gather all nodes $i$ with $x_i \geq x_{n+1}$ into a set and output it as the maximum independent set. Clearly, this reduction takes polynomial time.

  To show the correctness of this reduction, first note that if $k = n + 1$, the problem can be trivially solved with a solution by having $x_{n + 1} = 1$ and all other $x_i = 0$. This will result in a solution satisfying $nm$ inequalities and hence $t$ (the system origin) is among the top-$(n + 1)$ for such a weight vector. Of course, in this case, no nodes will be chosen for the maximum independent set. For a non-empty independent set with the size $a > 0$, we can assign $x_{n+1} = 1 / (a + 1)$, $x_i = 1/ (a + 1)$ for each node $i$ in the independent set and $x_i = 0$ otherwise. This will result in a vector $x$ satisfying $nm + a$ inequalities and hence, the Fair Top-$k$ Selection algorithm can be successfully solved with $k = n + 1 - a$ (recall our discussion in Section \ref{sec:hardness_fixed}). The maximum independent set has the largest $a$, which makes the value of $k$ smallest among all possible (non-empty) independent sets. This suggests that the maximum independent set can be obtained from the solution to the instance of the Fair Top-$k$ Selection problem with the smallest $k$.

  However, we have not ruled out the possibility that the weight vector obtained by solving the Fair Top-$k$ Selection problem with the smallest $k$ does not give us an independent set. Now, we show that for any $1 \leq k \leq n$, the solution must correspond to a (non-empty) independent set. Assume, for sake of contradiction, that the corresponding subset of nodes is not an independent set. Then we must have a pair of nodes $i$ and $j$ that $x_i \geq  x_{n+1}$, $x_j \geq x_{n+1}$ and $\{i, j\} \in E$. This results in a violation of the inequality $x_i + x_j - x_{n+1} \leq 0$ (with $x_{n + 1} \neq 0$). Since we have $n$ variants of this inequality, $n$ inequalities of the form $bx_i + bx_j - bx_{n+1} \leq 0$ are violated. In such a case, for the weight vector $x$, there are at least $n$ candidates whose scores are greater than that of $t$ (the system origin), $t$ cannot be among top-$k$ for any $1 \leq k \leq n$, and this is a contradiction.

  To conclude, the solution to the Fair Top-$k$ Selection problem with $1 \leq k \leq n$ will always yield an independent set, and the assignment of $x$ induced by the maximum independent set will satisfy the maximum number of inequalities among all possible (non-empty) independent sets. This establishes the correctness of our reduction.
\end{proof}

\begin{rmk}\label{rmk:np}
    The failure of a direct extension of Corollary \ref{col:hd} may appear to reveal a gap in the hardness barrier, as the reduction breaks down under the widely adopted assumption that weight vectors are non-negative. Moreover,
    the non-negativity (including weight vectors) arising from our problem setup (recall Section~\ref{sec:prel}) appears to open the door to efficient mathematical programming techniques. Specifically, the problem can be formulated as finding a feasible solution to a mixed-integer linear program (see Section \ref{alg:milp}), where solvers may exploit the non-negativity of variables and coefficients for efficiency (e.g., \cite{ahmed2018relaxations, takazawa2019approximation}). These techniques typically operate outside the decision-tree model implied in Corollary \ref{col:hd}, suggesting that the Fair Top-$k$ Selection problem may admit a more efficient algorithm. However, our NP-hardness reduction rules out this possibility for the worst case. Notably, the relaxation technique, a cornerstone of many efficient MILP solvers, also plays a key role in our NP-hardness proof.
\end{rmk}

\section{From hardness results to algorithm design}\label{sec:design}
As implied by the hardness insights (Remarks~\ref{rmk:3sum}--\ref{rmk:np}), our two-pronged solution (small $k$ vs. large $k$) is guided by the hardness results, and these implications also warn against some algorithm design strategies that are unlikely to yield efficient solutions. In this section, we discuss how hardness results guide our algorithm design. Also, we examine the case of small $k$ in details, which is critical to our two-pronged solution.

\subsection{Implications for algorithm design}\label{subsec:implications}
Our hardness results begin by considering restricted cases, suggesting the problem becomes hard to solve beyond two dimensions (Remark~\ref{rmk:3sum}) and demonstrating the curse of dimensionality (Remark~\ref{rmk:hd}). Via attempts to extend the hardness results to more general and realistic cases, we have the following critical insights for the algorithm design:

\begin{itemize}
    \item \textbf{Small $k$ Opportunity.} The reduction technique used in the Theorem \ref{thm:k} breaks down for a sufficiently small $k$ (Remark \ref{rmk:k}). This suggests the potential of a more efficient algorithm by explicitly leveraging $k$ in the design.
    \item \textbf{Large $k$ Practicality.} The hardness of the problem is robust for a large $k$, making theoretically efficient algorithms unlikely in the worst case. Therefore, designing efficient algorithms for large $k$ should focus on real-world performance rather than worst-case runtime guarantees.
    \item \textbf{Fairness Relaxation Pitfall.} Relaxing the fairness constraint may appear to make the problem easier, making an algorithm that starts with a loose constraint and gradually tightens it seems attractive. However, such relaxation does not reduce worst-case computational hardness (Remark~\ref{rmk:relaxed}), limiting its value.
    \item \textbf{Non-negativity Trap.} With the failure to extend Corollary~\ref{col:hd}, non-negativity of variables (weights) and most coefficients (scoring attribute values) appears to open the door to efficient mathematical programming techniques (Remark~\ref{rmk:np}). However, our NP-hardness result shows that this non-negativity is unlikely to yield runtime improvements, at least in the worst-case scenario.
\end{itemize}

Drawing on these insights that identify promising directions and warn against potentially ineffective ones, we arrive at a two-pronged solution: a theoretically grounded algorithm for a small $k$, and a practically efficient algorithm for a large $k$. However, the implication of Theorem \ref{thm:k} (Remark \ref{rmk:k}) only shows that breaking the lower bound is possible, it does not guarantee that it can be done. Next, we demonstrate that this lower bound is indeed breakable for a small $k$.

\subsection{The case of small $k$}\label{subsec:small_k}
To show that the lower bound can be broken in the case of small $k$, the idea is to design an algorithm whose run time is sensitive to $k$ and ensure that it runs fast when the value of $k$ is small. The Fair Top-$k$ Selection problem can be solved with algorithms given in \cite{asudeh2019designing}, which were designed for a more general fairness model. In 2-D, a direct application of the algorithm in \cite{asudeh2019designing} runs in $O(n^2(k + \log n))$ time but can be improved to $O(n^2\log n)$. In higher dimensions, the algorithm in \cite{asudeh2019designing} runs in $O(n^{2d+2})$ time in fixed dimensions if implemented using the same algorithm for linear programming \cite{seidel1991small} employed in this work. Since these algorithms target at a more general fairness model, they do not incorporate the value of $k$ in their design. While it may seem appealing to adapt these methods by making them sensitive to $k$, there are two barriers: \textbf{(1)} The algorithms implicitly employ a dimensional projection for the purpose of weight vector space navigation, which causes the loss of ranking information that makes it difficult to integrate the notion of top-$k$. \textbf{(2)} The algorithms' run times are lower bounded by $\Omega(n^2)$ in principle due to their dependence on pairwise ordering exchanges, which make them undesirable for large datasets. While our algorithm employs some similar computational geometry techniques used in \cite{asudeh2019designing}, it differs in its design principle as it aims at incorporating $k$ to begin with, which enables it to break established lower bounds for a sufficiently small $k$.

Before delving into the details of our algorithm, let us introduce essential computational geometry concepts it relies on. First, similar to what was done in \cite{asudeh2019designing}, a dual transformation is going to be employed which transforms points in $\mathbb{R}^d$ into ($d-1$)-dimensional hyperplanes in $\mathbb{R}^d$. However, since weight vectors in our problem are normalized by their $L_1$ norm, instead of $L_2$ norm, we use a different dual transformation given in \cite{cao2017k, xiao2023rkhit}.

Let $p = (p_1, p_2, \dots, p_d)$ be values of the scoring attributes in $\mathbb{R}^d$ of a candidate $c\in C$. We transform $p$ into a hyperplane $H(p)$ defined by
\begin{equation}
   (p_1 - p_d)x_1 + (p_2 - p_d)x_2 + \cdots +(p_{d-1} - p_d)x_{d-1} - x_d + p_d = 0,
\end{equation}
where $x_1, x_2, \cdots x_d$ are variables. Thus, 
\begin{equation}
x_d = p_1x_1 + p_2x_2 + \cdots + p_{d - 1}x_{d-1} + (1 - \sum_{i = 1}^{d-1}x_i)p_{d}.
\label{eq:xd}
\end{equation}

\begin{figure}[hbpt]
    \captionsetup[subfigure]{justification=centering}
    \centering
\begin{minipage}[b]{0.3\columnwidth}
    \centering
    \includegraphics[width=.7\linewidth]{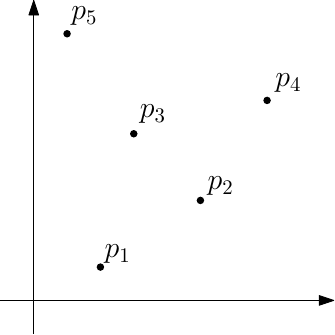}
\subcaption{Primal space}
\end{minipage}
\begin{minipage}[b]{0.3\columnwidth}
    \centering
    \includegraphics[width=.7\linewidth]{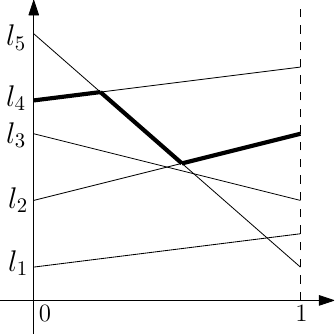}
    \subcaption{Dual space}\label{fig:dual2D_klevel}
\end{minipage}
\caption{Dual transformation in 2-D, mapping a point $p_i$ to a line $l_i$. Each (open) line segment is a cell in 2-D and bold line segments in (b) show the $k$-level for $k = 1$ in 2-D.}
\label{fig:dual2D}
\end{figure}

An example of this primal-to-dual transformation for a set of points in 2-D is provided in Figure~\ref{fig:dual2D}.

Recall that for any weight vector $w = (w_1, w_2, \dots, w_d)$ with $w_i \geq 0$ and $|w||_1 = 1$, we have $f(p) = \sum_{i=1}^d w_ip_i = \sum_{i=1}^{d - 1} p_iw_i + (1 - \sum_{i = 1}^{d-1}w_i)p_{d}$. Observe from Equation~(\ref{eq:xd}) that $f(p)$ is actually the $d$-th coordinate of the projection of point $(w_1,w_2, \ldots, w_{d-1})$ on $H(p)$. Now, consider two points $p$ and $p'$. If $f(p) > f(p')$ for some weight vector $w$, then in the dual space, the downward-directed ray from $(w_1,w_2, \ldots, w_{d-1},+\infty)$ hits $H(p)$ before $H(p')$.  The transformation thus preserves, in dual space, the relative order between each pair of points in primal space, with respect to their scores for any weight vector.  Notably, a tie in scores, where $f(p) = f(p')$, corresponds to an intersection between $H(p)$ and $H(p')$ in the dual space.

Next, consider the subdivision of $\mathbb{R}^d$ induced by a collection of hyperplanes, i.e., an \textit{arrangement}. We say that a point of $\mathbb{R}^d$ is at \textit{level~$k$} if exactly $k$ hyperplanes lie strictly above it in the positive $x_d$ direction. A \textit{cell} on a hyperplane is a maximal connected region of the hyperplane that is not intersected by other hyperplanes. The \textit{$k$-level} consists of the closure of all cells whose interior points have level $k$.  It is a monotone, piecewise-linear surface. (See Figure~\ref{fig:dual2D_klevel} for an example in 2-D.) The \textit{$\leq\! k$-level} is  the part of the arrangement lying on or above the $k$-level (see~\cite{halperin2017arrangements}). One important property of this $k$-level is that if a hyperplane has a cell that is part of the $(k-1)$-level, then the primal point corresponding to the hyperplane is the $k$th point for some weight vectors, since there are exactly $k-1$ points with higher scores. Moreover, if two weight vectors whose corresponding downward-directed rays in the dual space to hit the same cell of the $(k-1)$-level, their top-$k$ subsets are the same.

Having introduced all necessary concepts, let us proceed to the algorithm. Assume that we have an algorithm that constructs the $(k-1)$-level and identifies the cells adjacent to each cell of the $(k-1)$-level. For simplicity, we first assume that the hyperplanes are in general positions, i.e., no more than $d$ hyperplane intersect at a point, aligning with most theoretical algorithms for the $k$-level. Moreover, ties are broken arbitrarily when selecting the top-$k$ subset for a given a weight vector. Our algorithm first finds an arbitrary point, $t\in V$, where $V$ is described by the $l$ linear equalities (see Section~\ref{sec:prel}). Then, we find a cell $c$ of the $(k-1)$-level that is intersected by the downward-directed ray induced by $t$. Next, we obtain the corresponding top-$k$ subset of $t$ and count the number of points of $\mathcal{G}_1$ in the top-$k$ subset, denoted by $n_{k}^{\scriptscriptstyle \mathcal{G}_1}$, and test it against $L_{k}^{\scriptscriptstyle \mathcal{G}_1}$ and $U_{k}^{\scriptscriptstyle \mathcal{G}_1}$. If the top-$k$ subset satisfies the fairness constraint, we output the weight vector and stop. If not, we traverse the $(k-1)$-level by visiting adjacent cells in a depth-first manner. For each cell visited for the first time, we test if its projection onto the first $d - 1$ coordinates intersects $V$, If it does not, then we ignore the cell; otherwise, we update the top-$k$ subset as follows. Observe that the top-$k$ subsets of adjacent cells differ in at most one point since there is at most one hyperplane that enters the $\leq(k-1)$-level and becomes part of the $(k-1)$-level. We maintain the top-$k$ subset and $n_{k}^{\scriptscriptstyle \mathcal{G}_1}$ during the traversal and whenever the top-$k$ subset changes, test $n_{k}^{\scriptscriptstyle \mathcal{G}_1}$ against $L_{k}^{\scriptscriptstyle \mathcal{G}_1}$ and $U_{k}^{\scriptscriptstyle \mathcal{G}_1}$. The algorithm stops when either a fair weight vector is obtained from a $(k-1)$-level cell with $n_{k}^{\scriptscriptstyle \mathcal{G}_1}$ satisfying the fairness constraint or all cells of the $(k-1)$-level whose projections intersect $V$ have been visited.

For the runtime analysis, leveraging known results on the structural complexity of the $(k-1)$-level and existing algorithms for constructing it, along with a careful application of an efficient convex polyhedra intersection algorithm \cite{barba2014optimal}, we obtain the following:

\begin{thm}[Small-$k$ Tractability]\label{thm:k_level_runtime}
    In fixed dimensions, Fair Top-$k$ Selection can be solved in $O(nk^{1/3}\log n)$ time for $d=2$ and in $O(nk^{3/2}\log^4 n)$ time for $d=3$, both with $l = O(n)$. In $d\geq 4$ dimensions, it can be solved in $O(n^{\lfloor d/2 \rfloor}k^{\lceil d/2 \rceil})$ (expected) time with a sufficiently small $l$ (e.g., $l = O(\textnormal{\text{polylog}}(n))$).
\end{thm}

\begin{proof}
    The algorithm's run time depends on the $(k-1)$-level's structural complexity, as it visits all cells of the $(k-1)$-level whose projections intersect $V$. In the worst case, every cell of the $(k-1)$-level will be visited. Besides, the time for constructing the $(k-1)$-level should also be counted. Let $s$ be the number of cells of the $(k-1)$-level. The best known bounds for $s$ are $O(nk^{1/3})$ for $d=2$ \cite{dey1998improved}, $O(nk^{3/2})$ \cite{sharir2001improved} for $d=3$, and $O(n^{\lfloor d/2 \rfloor}k^{\lceil d/2 \rceil - c_d})$ for $d \geq 4$ \cite{agarwal1998levels} where $c_d$ is a small positive constant that depends on $d$. In 2-D, the $(k-1)$-level can be constructed in $O((n + s)\log n)$ time using the algorithm in \cite{edelsbrunner1986constructing} with the data structure in \cite{brodal2002dynamic}. In 3-D, the $(k-1)$-level can be constructed in  $O(n\log n+ s\log^4 n)$ time using the algorithm in \cite{agarwal1995dynamic} with the data structure in \cite{chan2020dynamic}. In higher dimensions ($d\geq 4$), the $(k-1)$-level can be constructed in $O(n^{\lfloor d/2 \rfloor}k^{\lceil d/2 \rceil})$ (expected) time \cite{mulmuley1991levels, agarwal1998constructing}. For each visited cell, additional time is spent for intersection and fairness tests. The fairness test adds $O(1)$ time per cell. For $d=2$, the intersection test adds $O(1)$ time per cell with $l = O(n)$. For higher dimensions ($d \geq 3$), it adds $O(\log l)$ time per cell by applying the technique in \cite{barba2014optimal}, with $l = O(n)$ for the $d = 3,\,4$ case, and with $l$ being sufficiently small (i.e., $l = O(n^{\frac{1}{\lfloor (d-1) / 2\rfloor} - \delta})$ for any constant $\delta > 0$) for $d > 4$.
\end{proof}

\begin{rmk}\label{rmk:k_level_runtime}
    For any $k = O(\textnormal{\text{polylog}}(n))$, the upper bounds become in $\tilde{O}(n)$ time for $d=2,\,3$, and more generally, $\tilde{O}(n^{\lfloor d/2 \rfloor})$ for any constant $d$ ($\tilde{O}(\cdot)$ hides a polylog $n$ factor). The algorithm breaks the $\Omega(n^{2-\delta})$ lower bound for $d=3$ and $\Omega(n^{d-1})$ lower bound for any fixed $d\geq3$ in case of small $k$. (Note that $k= O(\textnormal{\text{polylog}}(n))$ is not for choosing between algorithms in our two-pronged solution, see Section~\ref{subsec:choice}.) However, it is worth noting that the algorithm relies on assumptions (general position and arbitrary tie-breaking) that do not fully align with the problem setting (as discussed in Section~\ref{subsec:practical_klevel}). With careful refinements and thorough analysis, one can show that the bounds still hold. Nevertheless, since the main purpose of this algorithm is to demonstrate the key component for breaking the lower bounds, which informs the design of practical algorithms, its full resolution is addressed in that context (see Section~\ref{subsec:practical_klevel}).
\end{rmk}

\section{Practical algorithms}\label{sec:pract_algs}
The $k$-level-based algorithm presented in the previous section builds on theoretically efficient algorithms for constructing the $k$-level. However, there are no known empirical evaluations on most of them. Moreover, known empirical evaluations of a related algorithm indicated worse practical performance compared to simpler heuristics \cite{aharoni1999line, har2000taking}. This gap arises because a theoretical efficient algorithm does not always translate into real-world performance gains, due to discrepancies between theoretical models and practical realities. Moreover, many theoretical algorithms make certain assumptions that are not appropriate for our problem setting. In this section, we identify key engineering challenges that make direct application of such algorithms impractical and present solutions that overcome these limitations. Furthermore, as demonstrated in Theorem \ref{thm:k_level_runtime}, the run time of our $k$-level-based algorithm are sensitive to the value of $k$, especially when the dimensionality, $d$, is large. As noted in Section~\ref{subsec:implications}, for a large $k$ value, our algorithm design principle shifts from theoretical guarantees to practical efficiency. As a result, we present an algorithm based on mixed-integer linear programming (MILP) that, while being theoretically worse, performs better for real-world datasets. Since the practical performance becomes the focus, the choice between these two algorithms will be discussed in Section~\ref{subsec:choice} based on experimental results.

\subsection{Practical $k$-level-based algorithms}\label{subsec:practical_klevel}
As stated above, a direct implementation of the $k$-level-based algorithm may not work well in practice. In fact, there are three engineering challenges, which are as follows:
\begin{itemize}
    \item \textbf{Space Overhead.} The $k$-level-based algorithm constructs the entire $k$-level, whose structural complexity grows quickly with dataset size and dimensionality (Theorem~\ref{thm:k_level_runtime}). For large datasets, it may exhaust internal memory, requiring significantly slower external-memory solutions. Moreover, the user may seek a fair weight vector within a small, explainable region of the weight space, making a full $k$-level unnecessary. In such cases, an adapted version of existing algorithm \cite{asudeh2019designing} may perform better.
    \item \textbf{Hardware Utilization.} A prior theoretically efficient algorithm \cite{har2000taking} exhibits worse memory performance that contributes to its slower run time compared to simpler heuristics, especially when real-world data deviate from worst-case assumptions. Moreover, many algorithms are not naturally parallelizable and cannot exploit modern multi-core systems. Bridging the theory-practice gap for parallel algorithms is even more challenging \cite{blelloch2021spaa}.
    \item \textbf{Tie-breaking.} Most theoretical algorithms for constructing the $k$-level assume general position (limiting ties among candidates) and break ties arbitrarily for simplicity. However, for the fair top-$k$ selection, tie-breaking can affect the number of protected group candidates selected, impacting fairness.  In our problem setting, an algorithm must be designed to account for all possible tie-breaking outcomes when evaluating fairness (recall Section~\ref{sec:prel}).
\end{itemize}

All these challenges should be properly addressed for the $k$-level-based algorithm to work in practice. Next, we are going to present practical $k$-level-based algorithms, with particular attention to addressing these three challenges and discussing other critical engineering decisions. (We will revisit the issue of tie-breaking in the context of the MILP-based algorithm in Section~\ref{alg:milp}.)

\subsubsection{2-D $k$-level-based algorithm}\label{subsubsec:k_level_2d}
Our algorithm in 2-D is an adaptation of an algorithm in \cite{chan1999remarks}, which itself was also briefly outlined in \cite{basch1996reporting}. For completeness, we briefly describe its general idea; details can be found in that paper. The algorithm uses a vertical sweep-line, moving from left to right over the set of lines, to compute the $(k-1)$-level in 2-D. Specifically, at all times during the sweep, the algorithm tracks  the input line that has $k-1$ lines above it and $n-k$ lines below it. For this purpose,  two kinetic priority queues, $S_1$ and $S_2$ are used. $S_1$ (resp. $S_2$) is a min-priority (resp. max-priority) queue and it contains the $k$ (resp. $n-k$) lines with largest (resp. smallest) $y$-coordinate values at their intersection points with the sweep-line. Viewing the $x$-coordinate of the sweep-line as ``time'', the kinetic priority queue looks ahead for the time instant when the minimum (resp. maximum) element of $S_1$ (resp. $S_2$)  changes and updates the corresponding queue, where the  minimum (resp. maximum) element is the lowest  (resp. highest) line in the queue at the current sweep-line position. There are three events to consider during the sweep: The time when the minimum (resp. maximum) element of $S_1$ (resp. $S_2$) changes and the time when the minimum line in $S_1$ and the maximum line in $S_2$ cross. The algorithm processes  the earliest of these three events first and updates the data structures accordingly.

The kinetic priority queue uses the following operations for event processing: \textsc{Top()} to return the current minimum or maximum element (depending on the queue being queried). \textsc{\mbox{NextEventTime()}} to return the closest time instant (in the future) when the minimum (or maximum) of the queue changes or an internal data structure event for bookkeeping occurs. \textsc{Advance()} to advance time and update the minimum (or maximum) as needed. In particular, we add the \textsc{Repace()} operation to replace the  minimum (or maximum) of the queue, which is relevant for addressing one of the key challenges (explained later).

\begin{figure}[htbp]
    \captionsetup[subfigure]{justification=centering}
    \centering
\begin{subfigure}[b]{0.3\columnwidth}
    \centering
    \includegraphics[width=0.7\linewidth]{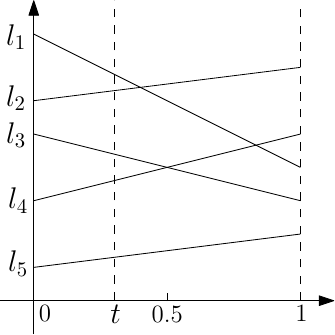}
\end{subfigure}
\begin{subfigure}[b]{0.3\columnwidth}
    \centering
    \resizebox{0.65\columnwidth}{!}{\begin{tikzpicture}[
      internal/.style={draw, rectangle split, rectangle split parts=2,align=center, rectangle split part align=center},
      leaf/.style={ draw, rectangle, align=center},
      level 1/.style={sibling distance=80pt},
      level 2/.style={sibling distance=40pt, level distance=50pt}, 
      level 3/.style={sibling distance=40pt, level distance=50pt},
      edge from parent/.style={draw, thick, -latex}
     ]
        \node[internal] {\LARGE $l_1$\nodepart{second}\Large 0.4}
    child {node[internal] {\LARGE $l_1$\nodepart{second} \Large 0.4}
        child {node[leaf] {\LARGE  $l_1$}}
        child {node[leaf] {\LARGE  $l_2$}}
    }
    child {node[internal] {\LARGE  $l_3$\nodepart{second}\Large 0.5}
      child {node[internal] {\LARGE  $l_3$\nodepart{second}\Large 0.5}
          child {node[leaf] {\LARGE  $l_3$}}
          child {node[leaf] {\LARGE  $l_4$}}
      }
      child {node[leaf] {\LARGE $l_5$}}
    };
    \end{tikzpicture}}
\end{subfigure}
\caption{Maximum kinetic tournament tree (right) for lines (left) at $x = t$. Additional auxiliary info. stored at internal nodes are omitted. For an internal node, the associated line (e.g, $l_1$ at the root) is the \textsc{Top()} for the (sub)tree rooted at that node at the given time (value of $x$), and the number is the closest time instant (in the future) for any nodes of the (sub)tree when the \textsc{Top()} of its left child intersects the \textsc{Top()} of its right child. We augment the tree with the \textsc{Replace()} operation, which can be implemented by substituting the target line (i.e., $l_1$ in this example) at the leaf level and propagating updates along the path to the root.}\label{fig:tourney}
\end{figure}

\subparagraph{Space overhead.} For reducing the space usage, notice that in this algorithm, we do not have to explicitly compute the $(k-1)$-level. The queue $S_1$ actually maintains the top-$k$ subset during the sweep. To find a fair weight vector, we only have to check whether the top-$k$ subset in $S_1$ satisfies the fairness constraint. Even better, following the theoretical algorithm (Section ~\ref{subsec:small_k}), we can use a variable to keep track of the number of lines (hence, indirectly, the primal points) in $S_1$ that belong to the protected group. Each time $S_1$ changes, we update this variable suitably and use it for the fairness test. As soon as the fairness test is passed, we stop the sweep and obtain a fair weight vector as given by the position of the sweep-line. Since explicit construction of the $(k-1)$-level is not needed, the space usage reduces to $O(n)$, eliminating the superlinear space overhead. Moreover, avoiding the explicit construction also improves runtime efficiency, yielding gains in both space and time.

\subparagraph{Hardware utilization.} Our low-level optimizations focus on the kinetic priority queue, a critical component for the algorithm's real-world performance, particularly for situations that divert from worst-case scenarios (see discussion in Section \ref{subsec:exp2d}). In \cite{chan1999remarks}, fully dynamic kinetic priority queues are used, which may result in frequent memory allocations/deallocations and bad memory access patterns. However, observe that the sizes of $S_1$ and $S_2$ are actually fixed throughout the sweep (at $k$ and $n-k$), kinetic priority queues do not need to be fully dynamic. To exchange two elements between the queues, we only have to replace the minimum (resp. maximum) element of $S_1$ (resp. $S_2$) with the maximum (resp.minimum) element of $S_2$ (resp. $S_1$) and update the queue accordingly. We use the kinetic tournament tree \cite{basch1996reporting} for implementing the kinetic priority queue. For the tournament tree, this replacement can be done in $O(\log n)$ time, and we implement an operation called \textsc{Replace()} for it (see Figure~\ref{fig:tourney}). Since the tree configuration is fixed (a full binary tree), we can further flatten trees into linear arrays for better cache performance. In our implementation, a pre-order traversal layout is used. (Other layouts might work better \cite{khuong2017array}, but are not explored in this work.)

\begin{figure}[htbp]
    \captionsetup[subfigure]{justification=centering}
    \centering
\begin{subfigure}[b]{0.3\columnwidth}
    \centering
    \includegraphics[width=0.7\linewidth]{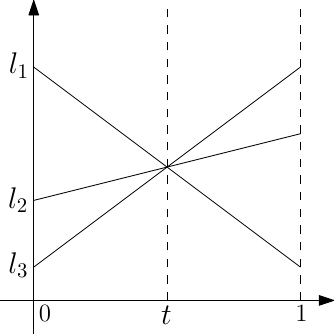}
    \subcaption{}\label{fig:tie1}
\end{subfigure}
\begin{subfigure}[b]{0.3\columnwidth}
    \centering
    \includegraphics[width=0.7\linewidth]{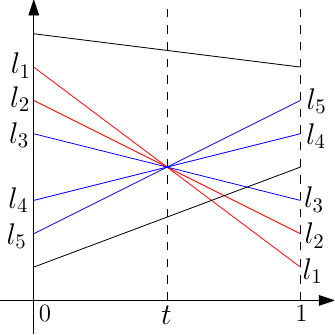}
    \subcaption{}\label{fig:tie2}
\end{subfigure}
\caption{\textbf{(a)} Three lines, $l_1 : y = m_1x + b_1$, $l_2 : y = m_2x + b_2$ and $l_3 : y = m_3x + b_3$ intersect at $x = t$. The tie is broken by perturbing $x$ to $x = t + \epsilon$, where $\epsilon$ is infinitesimally small and positive. This, in fact, sorts the lines by values of  $m_i$, and orders $l_3$ ahead of $l_1$ and $l_2$ for a max-queue. \textbf{(b)}~For $k = 3$, we have $l_1, l_2 \in S_1$ (red) and $l_3, l_4, l_5 \in S_2$ (blue) before the intersection. At $x = t + \epsilon$, we should have $l_5, l_4 \in S_1$ and $l_3, l_2, l_1 \in S_2$. Since $S_1.$\textsc{Top()} is $l_2$ and $S_2.$\textsc{Top()} is $l_3$ before the intersection, the algorithm finds $l_1$ in $S_1$ and $l_5$ in $S_2$ by repeatedly calling \textsc{Advance()}. The two lines are then exchanged. Now, we have $S_1.$\textsc{Top()} is $l_2$ and $S_2.$\textsc{Top()} is $l_4$ and the pair should also be exchanged.}
\end{figure}

\subparagraph{Tie-breaking.} The algorithm in \cite{chan1999remarks} assumes that the lines are in general position, i.e., no more than two lines intersect at a point. If not, then several events may occur simultaneously, and it is not clear if these simultaneous events can be processed in an arbitrary order without affecting the correctness of the algorithm. To resolve this issue, let us start by considering the case where the only events involved are of the first two types mentioned earlier. The issue becomes one of ensuring the correctness of the minimum  (or maximum) element as the sweep-line moves forward. This can be resolved via symbolic perturbation \cite{edelsbrunner1990simulation}, which determines the order of intersecting lines by comparing the slopes of lines if multiple lines intersect at a point. (See Figure~\ref{fig:tie1} for an example.) Symbolic perturbation ensures the correctness of the minimum (resp. maximum) element as the sweep-line passes through the intersection point. Note that applying symbolic perturbation here does not affect the tie-breaking for obtaining a top-$k$ subset, as the changes occur only within $S_1$ and $S_2$.

Next, consider the case where an event of the last type also occurs at the same time. Here, multiple elements could be exchanged between the two queues. We now need to determine which among these intersecting lines should be exchanged. (See Figure~\ref{fig:tie2} for an example.) We observe that the lowest (resp. highest) line at the intersection point (after the symbolic perturbation) is the highest (resp. lowest) line in $S_1$ (resp. $S_2$) right before the intersection. To obtain the lowest (resp. highest) line at the intersection point  in $S_1$ (resp. $S_2$), we keep calling \textsc{Advance()} until \textsc{\mbox{NextEventTime()}} is not the time instant for the intersection. Then, one can exchange this pair of lines. We keep doing this until $S_1.$\textsc{Top()} lies above $S_2.$\textsc{Top()} (after the symbolic perturbation). Note that the resulting $S_1$ should not be directly tested against the fairness constraint, as it is only one of the tie-breaking result. 

Having discussed how to handle simultaneous events, we turn to handle the tie-breaking for the top-$k$ subset itself. In 2-D, this issue originates from two sources: (1) Multiple lines intersect at one point and (2) Duplicate lines. To address them, each time an event of the last type is to be processed, we collect all intersecting lines at that time instant (including duplicates) by a guided tree traversal. Each intersecting line can be found with an additional $O(\log n)$ cost. Let $M_1$ (resp. $M_2$) be the set of intersecting lines in $S_1$ (resp. $S_2$) at time $t$. There are $k-|M_1|$ lines lying strictly above the $(k-1)$-level and for a valid top-$k$ subset, $|M_1|$ lines can be chosen from $M_1 \cup M_2$. To avoid the cost of enumerating all possible top-$k$ subsets, one can use a greedy method to compute tight lower and upper bounds on the number of lines of the protected group and use these bounds to test against the fairness constraint. For efficient calculation of these bounds, $S_1$ also maintains the number of elements in it that belong to $\mathcal{G}_1$, which can be obtained by calling \textsc{PGCount()}. Calculation of these bounds and testing fairness are done in the \textsc{IsFair()} function.

\begin{algorithm}[tb!]
    \caption{Practical 2-D $k$-level-based algorithm} \label{alg:klevelpract-2d}
    \begin{algorithmic}[1]
    \Procedure{KLevelBased2D}{$C$, $k$, $L_{k}^{\scriptscriptstyle \mathcal{G}_1}$, $U_{k}^{\scriptscriptstyle \mathcal{G}_1}$, $lb$, $ub$}
    \State Let $H$ be $C$ with each $p(c)$ transformed into a line.
    \State $H_1 \gets$ Top-$k$ subset of $H$ with $x = lb$.
    \State {\bf if} {$H_1$ satisfies the fairness constraint} {\bf then} \Return $(lb, 1- lb)$
    \State Build $S_1$ with $H_1$ and $S_2$ with $H \setminus H_1$.
    \MRepeat
    \State $t_1 \gets S_1.$\Call{NextEventTime}{}(); $t_2 \gets S_2.$\Call{NextEventTime}{}()
    \State $t_3 \gets$ \Call{IntersectTime}{}($S_1.$\Call{Top}{}(), $S_2.$\Call{Top}{}())
    \State $t \gets \min(t_1, t_2, t_3)$
    \State {\bf if} {$t > ub$} {\bf then} \Return NIL
    \If {$t = t_3$}
    \If {$t = t_1$}
    \Repeat {} $S_1.$\Call{Advance}{}()
    \Until {$S_1.$\Call{NextEventTime}{}() $\neq t$}
    \EndIf
    \If {$t = t_2$}
    \Repeat {} $S_2.$\Call{Advance}{}()
    \Until{$S_2.$\Call{NextEventTime}{}() $\neq t$}
    \EndIf
    \State $M_1 \gets$ lines in $S_1$ intersect $S_1.$\Call{Top}{}() at $x = t$
    \State $M_2 \gets$ lines in $S_2$ intersect $S_2.$\Call{Top}{}() at $x = t$
    \State $n_k^{\scriptscriptstyle \mathcal{G}_1} \gets$ $S_1.$\Call{PGCount}{}()
    \State {\bf if} {\Call{IsFair}{}($L_{k}^{\scriptscriptstyle \mathcal{G}_1}$, $U_{k}^{\scriptscriptstyle \mathcal{G}_1}$, $n_k^{\scriptscriptstyle \mathcal{G}_1}$, $M_1$, $M_2$)} {\bf then} \Return $(t, 1- t)$
    \While {$S_1.$\Call{Top}{}() lies below $S_2.$\Call{Top}{}()}
    \State $h_1 \gets$ $S_1.$\Call{Top}{}(); $h_2 \gets$ $S_2.$\Call{Top}{}()
    \State $S_1.$\Call{Replace}{$h_2$}; $S_2.$\Call{Replace}{$h_1$}
    \EndWhile
    \ElsIf {$t = t_1$} $S_1.$\Call{Advance}{}()
    \Else {} $S_2.$\Call{Advance}{}()
    \EndIf
    \EndRepeat
    \EndProcedure
\end{algorithmic}
\end{algorithm}

\subparagraph{Algorithm overview.} Algorithm~\ref{alg:klevelpract-2d} provides the pseudocode for the whole algorithm, where $lb$ and $ub$ represent $V$ in the dual space after processing.

\subsubsection{Multi-dimensional $k$-level-based algorithm}\label{subsubsec:k_level_hd}
Our algorithm in higher dimensions is an adaptation of an algorithm in \cite{andrzejak1999optimization}, which was also implemented in \cite{asudeh2022finding} for the rank-regret minimization problem. While this algorithm is not as efficient in run time as a known theoretical algorithm, it may be better at exploiting modern multi-core systems as suggested in \cite{andrzejak1999optimization}. However, no parallel implementation currently exists, as \cite{andrzejak1999optimization} focused primarily on theoretical aspects, and the implementation in \cite{asudeh2022finding} did not pursue this direction.  For completeness, we briefly describe its general idea in the context of Fair Top-$k$ Selection. Notice that in this problem, we are interested not in the actual structure of the $(k-1)$-level but rather in the top-$k$ subset implied by a cell of the $(k-1)$-level. Recall that adjacent cells  differ by at most one element in their  top-$k$ subset. So, starting at a cell with its corresponding top-$k$ subset, one is able to visit its adjacent cell implicitly by testing each potential top-$k$ subset, obtained by substituting one element of the top-$k$ subset with an element outside the top-$k$ subset, is a valid one or not. The test can be done using a linear program, i.e., testing whether there exists a hyperplane separating the potential top-$k$ subset from all others. In this way, one can enumerate all possible top-$k$ subsets without explicitly constructing the $k$-level via a method akin to breadth-first search. 

With the general idea of the algorithm, let us turn to how to address those three challenges.

\begin{figure}[htbp]
    \centering
    \resizebox{0.3\columnwidth}{!}{\begin{tikzpicture}
        \begin{axis}[y dir=reverse, xmin=0, xmax=1, ymin=0, ymax=1, zmin=0, zmax=1, 
            colormap = {slategraywhite}{rgb255=(220,220,220) rgb255=(180,180,180) rgb255=(140,140,140)}]
            \addplot3 [patch, patch type=triangle, draw=black, shader=flat, thick]
                coordinates { 
                    (0, 0, 0.8) (0.2, 0.0, 0.5) (0.0, 0.3, 0.65) 
                    (0.2, 0.0, 0.5) (0.0, 0.3, 0.65) (0.3, 0.3, 0.4)
                    (0.3, 0.3, 0.4) (0.2, 0.0, 0.5) (1.0, 0.0, 0.75)
                    (0.3, 0.3, 0.4) (0.0, 1.0, 0.85) (0.0, 0.3, 0.65)
                    (0.3, 0.3, 0.4) (0.6, 0.4, 0.42) (1.0, 0.0, 0.75)
                    (0.0, 1.0, 0.85) (0.4, 0.6, 0.42) (0.3, 0.3, 0.4)
                    (0.3, 0.3, 0.4)  (0.6, 0.4, 0.42) (0.4, 0.6, 0.42)
                };
            \addplot3[color=black, dashed, very thick]
                coordinates { (0.4, 0.15, 0.0) (0.2, 0.25, 0.0) (0.3, 0.6, 0.0)  (0.5, 0.4, 0.0)  (0.4, 0.15, 0.0)};
            
            \node at (axis cs:0.35, 0.35, 0.0) {\large $V$};
        \end{axis}
    \end{tikzpicture}}
\caption{$(k-1)$-level (triangular mesh) and $V$ (dashed region on the $x$-$y$ plane) in 3-D, as each weight vector $w$ corresponds to a point $(w_1, w_2)$ on the $x$-$y$ plane. Testing cell intersection is equivalent to finding a point in $V$ whose  downward-directed ray from $(w_1, w_2, +\infty)$ (Section ~\ref{subsec:small_k}) hits the cell (triangle). The constraints (points in $V$) can be incorporated into the linear program.}\label{fig:intersection}
\end{figure}

\subparagraph{Space overhead.} Since the structural complexity of the entire $k$-level grows more rapidly in higher dimensions, reducing space overhead becomes more critical. In the original algorithm, one uses a linear program to test whether a potential top-$k$ subset is valid or not. Given the constraints defining the weight vector space ($V$), the linear program can incorporate these constraints to determine whether a corresponding cell intersects $V$, by only considering weight vectors within $V$ when testing a potential top-$k$ subset (see Figure~\ref{fig:intersection}). This approach reduces the number of cells (top-$k$ subsets) visited during the navigation, and thus space usage, especially when $V$ only defines a small region of the weight space. Notably, this approach also improves the run time in such case, especially when $l$ (the number of linear inequalities) is small, as the overhead of incorporating and enforcing the constraints in the linear program remains low.

\subparagraph{Hardware utilization.} Since the algorithm may be parallelizable, our focus is to turn this potential into real-world performance gains. Notice that the algorithm can be regarded as a breadth-first search algorithm on a graph, where each cell is a node and each edge connects a pair of adjacent cells. Therefore, a method for parallel breadth-first search can be applied. However, since the graph is not known explicitly and finding an adjacent node can be expensive, efficient but sophisticated level-synchronous parallel algorithms such as the methods in \cite{leiserson2010work, shun2013ligra} may be difficult and less beneficial to adapt. Here we take a simpler approach by distributing jobs for finding adjacent nodes. Each thread independently takes a job for identifying an adjacent node and tests whether the given potential top-$k$ subset is valid or not. If it is, the thread generates subsequent jobs by enumerating all potential top-$k$ subsets from this top-$k$ subset. All threads terminate once a top-$k$ subset (with its corresponding weight vector) satisfying the given fairness constraint is found, or all top-$k$ subsets (i.e., $(k-1)$-level cells) have been processed. We use a global job queue for parallel job distribution. To reduce synchronization costs, since a strict FIFO order is not needed, a lock-free queue of a relaxed version of FIFO \cite{kirsch2013fast} is used (with a memory reclamation scheme \cite{brown2015reclaiming}). Besides, for testing whether a top-$k$ subset has already been processed, we use an implementation of the lock-free hash table \cite{shalev2006split}. Since the size of the table increases monotonically, we use an insertion-only table and disable memory reclamation, for further reducing synchronization costs and runtime overheads.

Moreover, our strategy of avoiding explicit lock usage for efficiency introduces some subtle implementation issues that have to be addressed, especially in deciding when to stop the algorithm. In our implementation, we use an atomic variable to keep track of the number of threads which does not find the job queue empty. To prevent premature thread termination, each thread increments the variable \textit{after} the linearization point \cite{herlihy1990linearizability} that the thread can decide the non-emptiness of the queue, but \textit{before} the linearization point that it pops an element out. Besides, it decrements the variable \textit{after} the linearization point (under the restricted out-of-order specification \cite{henzinger2013quantitative}) that it finds the queue empty. Even though this might result in occasional false positives for the existence of pending jobs, it eliminates the risk of premature thread termination, and these false positives can be quickly resolved. Similar issues are also handled for returning a single fair weight vector when multiple threads find ones simultaneously, particularly under a weak memory model~\cite{boehm2008foundations, batty2015problem}.

\subparagraph{Tie-breaking.} Handling ties in this algorithm can be relatively simple. For each linear program, instead of finding a hyperplane that strictly separates the top-$k$ subset, one can modify the linear program by allowing some points to lie on the hyperplane. This approach enables to algorithm to enumerate all possible top-$k$ subsets, which is justified by regarding a typical top-$k$ subset with ties as a lower dimensional cell formed by intersections of multiple hyperplanes; that is, a $j$-dimensional cell ($0 \leq j \leq d - 1$) corresponding a maximal connected region of the hyperplane of dimension $j$ in the intersection of a subset of the hyperplanes that is not intersected by any other hyperplane (see \cite{halperin2017arrangements} for a formal definition). In the extreme case (e.g., all hyperplanes intersect at a single point), the run time will increase by a lot. However, this rarely occurs in practice. Nevertheless, we still trade some efficiency for ease of implementation.

\subparagraph{Algorithm overview.} Algorithm~\ref{alg:klevelpract-hd} provides the pseudocode for the algorithm. Besides overcoming the above challenges, there are other engineering decisions relevant to its performance. For the choice of algorithm for linear programming, since the dimensionality is usually small in comparison to the size of the dataset, we use an implementation of Seidel's algorithm \cite{seidel1991small}, which runs in linear (expected) time in fixed dimensions and works well in practice when the dimensionality, $d$, is small (typically $d < 10$). Moreover, we ignore a potential top-$k$ subset if the element being swapped out dominates (recall Section \ref{sec:hardness_general}) the element being swapped in, since the former element always has a higher score for any weight vector.

\begin{algorithm}[tb!]
    \caption{Practical multi-dimensional $k$-level-based algorithm} \label{alg:klevelpract-hd}
    \begin{algorithmic}[1]
        \Procedure{KLevelBasedMD}{$C$, $k$, $L_{k}^{\scriptscriptstyle \mathcal{G}_1}$, $U_{k}^{\scriptscriptstyle \mathcal{G}_1}$, $V$}
        \State Let $Q$ be a parallel job queue
        \State Let $T$ be a parallel hash table storing top-$k$ subsets
        \State Find an arbitrary $w \in V$ and obtain a top-$k$ subset $\tau$
        \State {\bf if} \Call{IsFair}{$L_{k}^{\scriptscriptstyle \mathcal{G}_1}$, $U_{k}^{\scriptscriptstyle \mathcal{G}_1}$, $\tau$} {\bf then} \Return $w$ 
        \State $T \gets \{ \tau \}$
        \State \Call{EnumerateKSet}{$\tau$, $C$, $Q$, $T$}
        \ParWhile {one thread does not find $Q$ empty}
        \State $\tau \gets Q.$\Call{Dequeue}{}()
        \State {\bf if} $\tau=$ NIL {\bf then continue} \Comment{$Q$ is found empty}
        \State Find a $w \in V$ such that $\tau$ is a top-$k$ subset using a LP
        \If {the LP is feasible}
        \State {\bf if} \Call{IsFair}{$L_{k}^{\scriptscriptstyle \mathcal{G}_1}$, $U_{k}^{\scriptscriptstyle \mathcal{G}_1}$, $\tau$} {\bf then} \Return $w$
        \If {$\tau \notin T$}
        \State $T \gets T \cup \{\tau \}$
        \State \Call{EnumerateKSet}{$\tau$, $C$, $Q$, $T$}
        \EndIf
        \EndIf
        \EndParWhile
        \State \Return NIL
        \EndProcedure
    \end{algorithmic}

    \begin{algorithmic}[1]
    \Procedure{EnumerateKSet}{$\tau$, $C$, $Q$, $T$}
    \For {$o \in \tau$}
    \For {$c \in C \setminus \tau$}
    \State $\tau' \gets (\tau \setminus \{ o\}) \cup \{ c \}$
    \State {\bf if} $\tau' \notin T$ {\bf then} $Q.$\Call{Enqueue}{$\tau'$}
    \EndFor
    \EndFor
    \EndProcedure
    \end{algorithmic}
\end{algorithm}

\subsection{Mixed-integer linear programming-based algorithm}\label{alg:milp}
For a large $k$ value, our focus becomes finding a practically efficient algorithm, which results in an algorithm based on mixed-integer linear programming (MILP). We note that MILP is NP-hard and all known algorithms run in exponential times in the worst case. However, in practice, an MILP solver can be quite fast, and it was shown that an MILP-based algorithm is efficient for finding a linear scoring function for the reverse top-$k$ query \cite{chen2023not}. Experimental results show that our MILP-based algorithm can also be efficient for solving the Fair Top-$k$ Selection problem in practice.

To introduce our algorithm, for simplicity, assume without loss of generality that values of the scoring attributes are in the range of $[0,1]$ (see Section \ref{sec:prel}). For each candidate $c$, define a corresponding indicator variable $\delta_c \in \{0, 1\}$ satisfying
\begin{equation}\label{ieq:milp}
    -1 \leq w \cdot p(c) - \lambda - \delta_c \leq 0
\end{equation}
for any given weight vector $w$ and a cut-off value $\lambda\in [0, 1]$. To understand the inequalities, notice that a pair  $(w,\lambda)$ can be regarded as a hyperplane. For a top-$k$ subset $\tau_k^w$, there exists a $\lambda \in [0, 1]$ such that for any $c \in \tau_k^w$ and for any  $c' \in C \setminus\tau_k^w$, $w\cdot p(c) \geq \lambda$ and $w\cdot p(c') \leq \lambda$. The variable $\delta_c$ is used to encode whether or not a candidate $c$ is in the top-$k$ subset for a given $w$ and $\lambda$. More formally, we have
\begin{lem}\label{lem:milp}
    Consider a non-negative weight vector with $||w||_1 = 1$, and a cut-off value $\lambda\in [0,1]$. Any candidate $c$ with $p(c)\in [0, 1]^d$ satisfying inequalities (\ref{ieq:milp}) must have 
    \begin{align*}
        \delta_c = 
        \begin{cases}
            0, & \text{ if } w\cdot p(c) < \lambda, \\
            1, & \text{ if } w\cdot p(c) > \lambda.
        \end{cases}
    \end{align*}
    The value of $\delta_c$ can be either $0$ or $1$ if $w\cdot p(c) = \lambda$. 
\end{lem}
\begin{proof}
    Since $p(c)\in [0, 1]^d$, $w$ is non-negative and $||w||_1 = 1$, we have $0 \leq w\cdot p(c) \leq 1$. For any $\lambda\in [0,1]$, we have $-1 \leq w\cdot p(c) - \lambda \leq 1$. So, if $w\cdot p(c) < \lambda$, then $\delta_c$ must be $0$. Otherwise, if  $\delta_p = 1$, we would have $w\cdot p(c) - \lambda - \delta_p < -1$ that violates the inequalities (\ref{ieq:milp}). Similarly, if $w\cdot p(c) > \lambda$, $\delta_c$ must be $1$. When $w\cdot p(c) = \lambda$, $w\cdot p(c) - \lambda - \delta_p$ will be $0$ or $-1$ for the two choices of $\delta_c$, respectively, and both of them satisfy inequalities (\ref{ieq:milp}).
\end{proof}

Note that Lemma~\ref{lem:milp} does not restrict the value of $\lambda$, which is also necessary to ensure $\delta_c$ encodes the top-$k$ subset membership. The restriction of $\lambda$ is actually applied by a constraint on indicator variables, that is, $\sum_c \delta_c = k$. With this constraint, a value of $\lambda$ that does not result in exactly $k$ candidates with $\delta_c = 1$ will not be feasible. Moreover, the indicator variables $\delta_c$ can be used to encode the fairness constraint as follows:
\begin{equation}\label{inq:milp}
    L_{k}^{\scriptscriptstyle \mathcal{G}_1} \leq \sum_{A(c) = \mathcal{G}_1} \delta_c \leq U_{k}^{\scriptscriptstyle \mathcal{G}_1}.
\end{equation}
Finally, constraints on the weight vector can also be easily incorporated. The resulting formulation is a MILP with a weight vector, a cut-off value, and $n$ indicator variables as its unknown variables. Finding a fair weight vector becomes finding a feasible solution to the MILP.

Now, let us revisit the "Non-negativity Trap" (see Section~\ref{subsec:implications}) and consider the approach that exploits the non-negativity of variables and coefficients to solve the MILP efficiently. In our MILP formulation, all variables ($w$, $\lambda$ and $\delta_c$) are non-negative and most of the coefficients ($p(c)$) are positive (by inequalities (\ref{ieq:milp})). Coefficients of $\lambda$ and $\delta_c$ are negative but those of $\delta_c$ can be more flexible. The main issue is the negative coefficient of $\lambda$, which prevent direct adaptations of the techniques such as those used in \cite{ahmed2018relaxations, takazawa2019approximation}. While one might hope to circumvent this inconvenience using other techniques, our NP-hardness result (Theorem~\ref{thm:np}) show that the problem is fundamentally intractable in the worst case, indicating that such techniques will not be effective in general. This also suggests that the practical success of the MILP-based algorithm is largely heuristic.

For our implementation, we use Gurobi \cite{gurobi}---a state-of-the-art, mathematical programming tool---to solve the MILP. Parallel algorithms can be utilized to speed-up  Gurobi for this purpose and they are also used in our experiments. Moreover, since the objective function is not required for the resulting MILP, the Gurobi solver is configured to prioritize finding a feasible solution. 

\subparagraph{Tie-breaking.} As discussed in engineering challenges of Section~\ref{subsec:practical_klevel}, the tie-breaking issue cannot be ignored in the MILP-based algorithm either. This consideration has actual been embodied in our MILP formulation which naturally addresses it. Notice that if $w\cdot p(c) = \lambda$, $\delta_c$ is free to choose its value. So if the top-$k$ subset is not unique for a weight vector $w$, all possible top-$k$ subsets will be considered in the MILP since values of $\delta_c$ resulting from a top-$k$ subset are all feasible ones if there is no fairness constraint. The fairness constraint is applied to further restrict values of $\delta_c$ such that the top-$k$ subset derived from these values is fair.

\section{Experiments}
\label{sec:expts}
In this section, we present experimental results for our algorithms in 2-D and higher dimensions on real-world datasets.
We evaluate performance in terms of runtime and outcome quality, and conclude with key observations and discussions on how to choose between the $k$-level-based and MILP-based algorithms.
\subsection{Experimental setup}
\subparagraph{Experiments Design.} Although the definition of the problem allows for more flexibility, in our experiments, a weight vector $w^o$ was provided as an input. If $w^o$ was unfair, the algorithm was run to find a fair weight vector $w^{\scriptscriptstyle f}$ that does not vary from the original one by a lot. More formally, for a given unfair weight vector $w^o$, one aims to find a fair weight vector $w^{\scriptscriptstyle f}$ such that $|w^{\scriptscriptstyle f}_i - w^o_i| \leq \epsilon$ for all $ 1 \leq i \leq d$, with $\epsilon$ is user-specified parameter that limits the allowable deviation. In this setting, one might fail to find a fair weight vector that is close enough and the algorithm will simply report the failure and stop. We conducted two set of experiments, one to compare runtime performance and another to evaluate the quality of the algorithm's output.  Note that different methods were used to generate input weight vectors for these two sets of experiments. For the runtime comparison, weight vectors were sampled uniformly at random until $20$ \textit{unfair} weight vectors were found, and the run time was averaged over the same set of $20$ samples (excluding the time for sampling). A limit of 20 hours was set on the overall run time of an algorithm. For the outcome quality evaluation, $50$ weight vectors were sampled uniformly at random and used as inputs, regardless of whether they were fair or not. We omitted space usage experiments for $k$-level-based algorithms, as the proposed optimizations are not only essential for large datasets but can also improve the runtime efficiency by directly reducing the overall computational cost (see Sections~\ref{subsubsec:k_level_2d} and \ref{subsubsec:k_level_hd}).

\subparagraph{Datasets.} Two real-world datasets were used in our experiments (synthetic datasets will be discussed shortly). Details are given as follows:
\begin{itemize}
    \item \textbf{\textit{COMPAS}} \cite{compas}: This dataset consists of 7,214 defendants from Broward County in Florida between 2013 and 2014. It is collected and published by ProPublica for their investigation of racial bias in the criminal risk assessment software. {\tt juv\_other\_count}, {\tt c\_days\_from\_compas}, {\tt priors\_count}, {\tt start}, {\tt end} and {\tt c\_jail\_out} $-$ {\tt c\_jail\_in} (the number of days staying in jail) were used as scoring attributes (6-D), and {\tt juv\_other\_count} and {\tt c\_days\_from\_compas} were used for 2-D experiments. ``African-American'' ($51.2\%$ in the entire dataset) is the protected group.
    \item \textbf{\textit{IIT-JEE}} \cite{iitjee}: This dataset consists of scores of 384,977 students in the Mathematics, Physics, and Chemistry sections of IIT-JEE 2009 along with their genders and other personal info. All subject scores were used as scoring attributes (3-D), and the Physics and Chemistry scores were used for 2-D experiments. ``Female'' ($25.5\%$ in the entire dataset) is the protected group.
\end{itemize}

Before running an algorithm, all values of scoring attributes were normalized to the range of $[0, 1]$ (recall Section~\ref{sec:prel}). For our multi-dimensional experiments ($d \geq 3$), datasets were further preprocessed by shrinking the input size. For experiments in 3-D (IIT-JEE), the $k$-skyband \cite{papadias2005progressive} was obtained from the full dataset. The $k$-skyband of a point set includes all points that are dominated by at most $k - 1$ other points, and it is easy to verify that points in top-$k$ subsets of any weight vector must be fully contained in the $k$-skyband. For a fixed dimension, the $k$-skyband can be computed in $\tilde{O}(n)$ time \cite{sheng2012worst}. For experiments in 6-D (COMPAS), the size of $k$-skyband is relatively large due to increased dimensionality so the algorithm in \cite{chen2023not} was used to remove points that are not among top-$k$ for any weight vector. Since the size of the IIT-JEE dataset is larger, we used larger values of $k$ for it (ranging from $50$ to $500$), while smaller values of $k$ were used for the COMPAS dataset (ranging from $10$ to $100$).

\textbf{\textit{Note on synthetic datasets.}} Synthetic datasets were deliberately excluded from our experiments due to the problem's reliance on the bias against the protected group that distorts the data distribution. When scoring attributes are drawn from identical distributions for all groups, the need for a Fair Top-$k$ Selection algorithm diminishes. Generating a synthetic dataset which ``meaningfully'' reflects the bias, which often stems from social and historical factors, is beyond the scope of this work.

\subparagraph{Baselines.} For comparison, \textsc{2draysweep} and \textsc{ATC\textsubscript{$+$}} from \cite{asudeh2019designing} were implemented as baselines, with \textsc{2draysweep} for 2-D experiments and \textsc{ATC\textsubscript{+}} for multi-dimensional experiments (see Section~\ref{subsec:small_k} for their key ideas).  Since these baseline algorithms were designed for a more general fairness model, the following improvements were made to enhance their performance for the Fair Top-$k$ Selection problem:
\begin{itemize}
    \item \textsc{2draysweep}: Instead of visiting all top-$k$ candidates for the fairness test and doing it whenever the order of two candidates changes, the number of candidates of the protected group in the top-$k$ subset is maintained throughout the algorithm, and the fairness test is done only when there is an ordering exchange between a candidate in the top-$k$ subset and a candidate outside of it.
    \item \textsc{ATC\textsubscript{$+$}}: A hyperplane formed by an ordering exchange will be tested to see whether it intersects $V$ before inserting it into the arrangement tree, and it will be discarded if there is no intersection. 
\end{itemize}
Moreover, for the tie-breaking that was not thoroughly discussed in \cite{asudeh2019designing}, our implementation of \textsc{2draysweep} addresses it by accounting all tied candidates in the fairness test. Candidates with identical scoring attribute values are pre-identified to accelerate the test. However, it is not obvious how to handle tie-breaking in \textsc{ATC\textsubscript{$+$}} without introducing significant runtime overhead, so it is ignored in our implementation (\textsc{ATC\textsubscript{$+$}} only). Finally, despite our two algorithms for multi-dimensional experiments leveraging parallelism, a sequential implementation of \textsc{ATC\textsubscript{$+$}} was used in our experiments, as no parallel algorithms are known for building an arrangement tree. While parallel algorithms exist for building other geometric data structures \cite{sun2019parallel}, it is not obvious how to adapt these methods for building an arrangement tree (or an analogous binary space partition tree \cite{deberg-etal2008}). 

\subparagraph{Environment and Implementation.} All our implementations were in C\texttt{++} on a 128-core machine with 2x AMD EPYC 7763 (2.45 GHz) and 512GB RAM, and our codes\footnote{Available at \url{https://github.com/caiguangya/fair-topk}.\label{repo}} were compiled and run inside an Apptainer \cite{kurtzer2017singularity} container\footref{repo} with a Rocky Linux 8.10 host OS. Parallel algorithms were manually bound to one of the two CPUs (using taskset) if they used $64$ or fewer threads.

\subparagraph{Default values.} For experiments in 2-D, $k = 50$ and $\epsilon = 0.1$. For experiments in higher dimensions ($d \geq 3$), $k = 50$ and $\epsilon = 0.05$. For the fairness constraint, the proportion of the protected group candidates in top-$k$ selection results was constrained to be between $40\%$ and $60\%$ for COMPAS and between $10\%$ and $40\%$ for IIT-JEE. For parallelism, 64 threads were used.

\subsection{Runtime experiments}\label{subsec:expruntime}
Since the main focus of this work is the speed of execution, we first present the results of the runtime experiments, starting with the 2-D experiments followed by results in higher dimensions. We then demonstrate how better utilization of hardware enhances the performance of the $k$-level-based algorithms, which also provides additional context for more informed runtime comparisons.

\subsubsection{2-D runtime experiments}\label{subsec:exp2d}
This section presents and discusses the results of runtime experiments in 2-D. Notably, we discuss situations where the real-world datasets divert from the worst case scenario as shown in experimental results, and analyze the performance of our practical $k$-level algorithm beyond worst-case runtime analysis.

\begin{figure}[htbp]
    \captionsetup[subfigure]{justification=centering}
    \centering
\begin{minipage}[b]{0.3\columnwidth}
    \resizebox{\columnwidth}{!}{\begin{tikzpicture}
    \begin{axis}[
        width =\axisdefaultwidth,
        height= 180pt,
        legend style={font=\normalsize},
        legend pos=south east,
        xlabel={\LARGE $k$},
        ylabel={\Large Time (s)},
        xtick=data,
        xticklabels={10, 20, 50, 70, 100},
        xticklabel style = {font=\Large},
        yticklabel style = {font=\Large},
        mark size=4pt,
        ymode=log,
        ymin=1e-4,
        ymax=4e-3]
        \addplot[mark=o, color=blue] table [x=k, y=k-level-based, col sep=comma] {compas-2d-k.csv};
        \addplot[mark=diamond, color=black] table [x=k, y=baseline, col sep=comma] {compas-2d-k.csv};
        \legend{$k$-level-based, Baseline}
    \end{axis}
\end{tikzpicture}}
\subcaption{COMPAS, varying $k$}\label{2d-k1}
\end{minipage}
\hspace{10pt}
\begin{minipage}[b]{0.3\columnwidth}
    \resizebox{\columnwidth}{!}{\begin{tikzpicture}
        \begin{axis}[
            width =\axisdefaultwidth,
            height= 180pt,
            legend style={font=\normalsize, at={(0.54, 0.5)},anchor=north west},
            xlabel={\LARGE $k$},
            ylabel={\Large Time (s)},
            xtick=data,
            xticklabels={50, 70, 100, 200, 500},
            xticklabel style = {font=\Large},
            yticklabel style = {font=\Large},
            mark size=4pt,
            ymode=log,
            ymin=0.01,
            ymax=1]
            \addplot[mark=o, color=blue] table [x=k, y=k-level-based, col sep=comma] {jee-2d-k.csv};
            \addplot[mark=diamond, color=black] table [x=k, y=baseline, col sep=comma] {jee-2d-k.csv};
            \legend{$k$-level-based, Baseline}
        \end{axis}
    \end{tikzpicture}}
    \subcaption{IIT-JEE, varying $k$}\label{2d-k2}
\end{minipage}
\hspace{10pt}
\begin{minipage}[b]{0.3\columnwidth}
    \resizebox{\columnwidth}{!}{\begin{tikzpicture}
        \begin{axis}[
            width =\axisdefaultwidth,
            height= 180pt,
            legend style={font=\normalsize},
            legend pos=south east,
            xlabel={\LARGE $n$ (ratio of full size)},
            ylabel={\Large Time (s)},
            xtick=data,
            xticklabels={0.2, 0.4, 0.6, 0.8, 1.0},
            xticklabel style = {font=\Large},
            yticklabel style = {font=\Large},
            mark size=4pt,
            ymode=log,
            ymin=0.001,
            ymax=1.0]
            \addplot[mark=o, color=blue] table [x=epsilon, y=k-level-based, col sep=comma] {jee-2d-n.csv};
            \addplot[mark=diamond, color=black] table [x=epsilon, y=baseline, col sep=comma] {jee-2d-n.csv};
            \legend{$k$-level-based, Baseline}
        \end{axis}
    \end{tikzpicture}}
    \subcaption{IIT-JEE, varying $n$}\label{2d-n}
\end{minipage}
\vspace{4pt}
\newline
\begin{minipage}[b]{0.3\columnwidth}
    \resizebox{\columnwidth}{!}{\begin{tikzpicture}
    \begin{axis}[
        width =\axisdefaultwidth,
        height= 180pt,
        legend style={font=\normalsize},
        legend pos=south east,
        xlabel={\LARGE $\epsilon$},
        ylabel={\Large Time (s)},
        xtick=data,
        xticklabels={0.01, 0.05, 0.1, 0.15, 0.2},
        xticklabel style = {font=\Large},
        yticklabel style = {font=\Large},
        mark size=4pt,
        ymode=log,
        ymin=1e-4,
        ymax=4e-3]
        \addplot[mark=o, color=blue] table [x=epsilon, y=k-level-based, col sep=comma] {compas-2d-eps.csv};
        \addplot[mark=diamond, color=black] table [x=epsilon, y=baseline, col sep=comma] {compas-2d-eps.csv};
        \legend{$k$-level-based, Baseline}
    \end{axis}
\end{tikzpicture}}
\subcaption{COMPAS, varying $\epsilon$}\label{2d-eps1}
\end{minipage}
\hspace{10pt}
\begin{minipage}[b]{0.3\columnwidth}
    \resizebox{\columnwidth}{!}{\begin{tikzpicture}
        \begin{axis}[
            width =\axisdefaultwidth,
            height= 180pt,
            legend style={font=\normalsize, at={(0.54, 0.5)},anchor=north west},
            xlabel={\LARGE $\epsilon$},
            ylabel={\Large Time (s)},
            xtick=data,
            xticklabels={0.01, 0.05, 0.1, 0.15, 0.2},
            xticklabel style = {font=\Large},
            yticklabel style = {font=\Large},
            mark size=4pt,
            ymode=log,
            ymin=0.009,
            ymax=1.9]
            \addplot[mark=o, color=blue] table [x=epsilon, y=k-level-based, col sep=comma] {jee-2d-eps.csv};
            \addplot[mark=diamond, color=black] table [x=epsilon, y=baseline, col sep=comma] {jee-2d-eps.csv};
            \legend{$k$-level-based, Baseline}
        \end{axis}
    \end{tikzpicture}}
    \subcaption{IIT-JEE, varying $\epsilon$}\label{2d-eps2}
\end{minipage}
\caption{Runtime experimental results for 2-D datasets.}
\end{figure}

\subparagraph{Varying $\boldsymbol{k}$.} Figures~\ref{2d-k1} and \ref{2d-k2} show the average run times as $k$ varies. Our $k$-level-based algorithm performed consistently better than the baseline algorithm for all datasets. In the smaller one (COMPAS), the advantage of our $k$-level-based algorithm was relatively small but in the larger one (IIT-JEE), the advantage was significant---showing a speedup of over $40$x. One interesting finding is that the performance of our algorithm remained nearly unchanged for increasing $k$ values, which seems to contradict our theoretical analysis. In fact, the time spent on the line-sweeping process---originally intended for constructing the $k$-level---accounted for only a small fraction of the total run time in our experiments. The algorithm \cite{chan1999remarks} we adapted is output-sensitive, meaning it runs faster when the input lines have lower $k$-level structural complexity. It is known that the best known structural complexity of the $k$-level in 2-D, i.e., $O(nk^{1/3})$~\cite{dey1998improved}, might not be a tight upper bound \cite{toth2001point, halperin2017arrangements}, and the actual structural complexity of a variety of input distributions can be significantly lower \cite{chiu2020average}. Of course, given the range of $k$ ($k \leq 500$) in our experiments, the increment of $k$ may not increase the structural complexity of the $k$-level by a lot, even in the worst case. On the other hand, the performance of the baseline also stayed almost the same, which was expected since its run time is independent of $k$.

\subparagraph{Varying $\boldsymbol{\epsilon}$.} Figures~\ref{2d-eps1} and \ref{2d-eps2} show the average run times as $\epsilon$ varies. Our $k$-level-based algorithm still outperformed the baseline in all cases. The run time of the baseline algorithm increased with larger values of $\epsilon$, as this led to more ordering exchanges being processed by the algorithm. However, the performance of our $k$-level-based algorithm only slightly increased, which again reflects that the actual structural complexity of the $k$-level for a real-world dataset can be low.

\subparagraph{Varying $\boldsymbol{n}$.} Figure~\ref{2d-n} shows average run times as $n$ varies. We controlled $n$ by randomly selecting $20\%$, $40\%$, $60\%$, $80\%$, and $100\%$ of the IIT-JEE dataset. For the baseline algorithm, the run time increased mildly as $n$ increased, performing better in practice than its worst case time complexity would suggest. However, our algorithm still managed to outperform the baseline algorithm by a lot in all cases. Comparing the baseline algorithm and our algorithm side-by-side, one can see that both algorithms solve the problem by exchanging the order of pairs of lines as the sweep-line moves forward. The difference is that the baseline algorithm considers the ordering exchanges of all pairs of lines, and our $k$-level-based algorithm uses kinetic tournament trees to efficiently skip many line exchanges that will not affect the top-$k$ subset. In fact, our algorithm improves the run time by directly reducing the number of line exchanges.  In this context, it is crucial that operations on the kinetic tournament tree are as efficient as possible, since many still incur an  $O(\log n)$ overhead. The experiments reveal an area where an efficient implementation of our $k$-level-based algorithm should be focus, and our optimized implementation of kinetic tournament tree (see Section \ref{subsubsec:k_level_2d}) addresses this need effectively (see Section \ref{subsubsec:ablation}), enhancing the algorithm's performance in scenarios where the worst-case behavior does not manifest.

\subsubsection{Multi-dimensional runtime experiments}
This section presents and analyzes the results of multi-dimensional runtime experiments. Notably, we compare the performance of the $k$-level-based algorithm and the MILP-based algorithm under various experimental settings, providing insights to guide the choice between the two algorithms in subsequent discussions.

\begin{figure}[tb!]
    \captionsetup[subfigure]{justification=centering}
    \centering
\begin{minipage}[b]{0.3\columnwidth}
    \resizebox{\columnwidth}{!}{\begin{tikzpicture}
        \begin{axis}[
            width =\axisdefaultwidth,
            height= 180pt,
            legend style={font=\small, at={(0.77, 0.97)},anchor=north east},
            xtick=data,
            symbolic x coords={COMPAS, IIT-JEE},
            xticklabels={\Large COMPAS \\\Large{($k = 10 \text {, }\epsilon = 0.001$)}, \Large IIT-JEE \\\Large{($k = 50\text {, }\epsilon = 0.01$)}},
            x tick label style={
            align=center,
            },
            enlarge x limits= 0.5,
            bar width=13,
            ylabel={\Large Time (s)},
            yticklabel style = {font=\Large},
            log origin=infty,
            ybar,
            ymin=1e-3,
            ytick={1e-3, 1e-2, 1e-1, 1, 10, 100, 1000, 10000},
            ymode=log]
            \addplot[red, pattern=north east lines, pattern color=red] coordinates {(COMPAS, 2.010e-02)  (IIT-JEE, 2.470e-02)};
            \addplot[blue, pattern=north west lines, pattern color=blue] coordinates {(COMPAS, 6.6500e-03) (IIT-JEE, 2.105e-02)};
            \addplot [pattern=crosshatch] coordinates {(COMPAS, 2.304e+03)  (IIT-JEE, 8.887e+01)};
            \legend{MILP-based, $k$-level-based, Baseline}
        \end{axis}
\end{tikzpicture}}
\subcaption{Baseline comparison}\label{md-baseline-compare}
\end{minipage}
\hspace{10pt}
\begin{minipage}[b]{0.3\columnwidth}
    \resizebox{\columnwidth}{!}{\begin{tikzpicture}
    \begin{axis}[
        width =\axisdefaultwidth,
        height= 180pt,
        legend style={font=\small},
        legend pos=south east,
        xlabel={\LARGE $k$},
        ylabel={\Large Time (s)},
        xtick=data,
        xticklabels={10, 20, 50, 70, 100},
        yticklabel style = {font=\Large},
        xticklabel style = {font=\Large},
        mark size=4pt,
        ymode=log]
        \addplot[mark=square, color=red] table [x=k, y=milp-based, col sep=comma] {compas-k.csv};
        \addplot[mark=o, color=blue] table [x=k, y=k-level-based, col sep=comma] {compas-k.csv};
        \legend{MILP-based, $k$-level-based}
    \end{axis}
\end{tikzpicture}}
\subcaption{COMPAS, varying $k$}\label{hd-k1}
\end{minipage}
\hspace{10pt}
\begin{minipage}[b]{0.3\columnwidth}
    \resizebox{\columnwidth}{!}{\begin{tikzpicture}
        \begin{axis}[
            width =\axisdefaultwidth,
            height= 180pt,
            legend style={font=\small},
            legend pos=south east,
            xlabel={\LARGE $k$},
            ylabel={\Large Time (s)},
            xtick=data,
            xticklabels={50, 70, 100, 200, 500},
            yticklabel style = {font=\Large},
            xticklabel style = {font=\Large},
            mark size=4pt,
            ymode=log]
            \addplot[mark=square, color=red] table [x=k, y=milp-based, col sep=comma] {jee-k.csv};
            \addplot[mark=o, color=blue] table [x=k, y=k-level-based, col sep=comma] {jee-k.csv};
            \legend{MILP-based, $k$-level-based}
        \end{axis}
    \end{tikzpicture}}
    \subcaption{IIT-JEE, varying $k$}\label{hd-k2}
\end{minipage}
\vspace{4pt}
\newline
\begin{minipage}[b]{0.3\columnwidth}
    \resizebox{\columnwidth}{!}{\begin{tikzpicture}
        \begin{axis}[
            width =\axisdefaultwidth,
            height= 180pt,
            legend style={font=\small},
            legend pos=north west,
            xlabel={\LARGE $n$ (ratio of full size)},
            ylabel={\Large Time (s)},
            xtick=data,
            xticklabels={0.2, 0.4, 0.6, 0.8, 1.0},
            yticklabel style = {font=\Large},
            xticklabel style = {font=\Large},
            mark size=4pt,
            ymode=log]
            \addplot[mark=square, color=red] table [x=n, y=milp-based, col sep=comma] {compas-n.csv};
            \addplot[mark=o, color=blue] table [x=n, y=k-level-based, col sep=comma] {compas-n.csv};
            \legend{MILP-based, $k$-level-based}
        \end{axis}
    \end{tikzpicture}}
    \subcaption{COMPAS, varying $n$}\label{hd-n1}
\end{minipage}
\hspace{10pt}
\begin{minipage}[b]{0.3\columnwidth}
    \resizebox{\columnwidth}{!}{\begin{tikzpicture}
        \begin{axis}[
            width =\axisdefaultwidth,
            height= 180pt,
            legend style={font=\small},
            legend pos=south east,
            xlabel={\LARGE $n$ (ratio of full size)},
            ylabel={\Large Time (s)},
            xtick=data,
            xticklabels={0.2, 0.4, 0.6, 0.8, 1.0},
            yticklabel style = {font=\Large},
            xticklabel style = {font=\Large},
            mark size=4pt,
            ymode=log,
            ymin=0.05,
            ymax=1.0]
            \addplot[mark=square, color=red] table [x=n, y=milp-based, col sep=comma] {jee-n.csv};
            \addplot[mark=o, color=blue] table [x=n, y=k-level-based, col sep=comma] {jee-n.csv};
            \legend{MILP-based, $k$-level-based}
        \end{axis}
    \end{tikzpicture}}
\subcaption{IIT-JEE, varying $n$}\label{hd-n2}
\end{minipage}
\hspace{10pt}
\begin{minipage}[b]{0.3\columnwidth}
    \resizebox{\columnwidth}{!}{\begin{tikzpicture}
        \begin{axis}[
            width =\axisdefaultwidth,
            height= 180pt,
            legend style={font=\small},
            legend pos=south east,
            xlabel={\LARGE $d$},
            ylabel={\Large Time (s)},
            xtick=data,
            xticklabels={3, 4, 5, 6},
            yticklabel style = {font=\Large},
            xticklabel style = {font=\Large},
            mark size=4pt,
            ymode=log]
            \addplot[mark=square, color=red] table [x=d, y=milp-based, col sep=comma] {compas-d.csv};
            \addplot[mark=o, color=blue] table [x=d, y=k-level-based, col sep=comma] {compas-d.csv};
            \legend{MILP-based, $k$-level-based}
        \end{axis}
    \end{tikzpicture}}
\subcaption{COMPAS, varying $d$}\label{hd-d}
\end{minipage}
\vspace{4pt}
\newline
\begin{minipage}[b]{0.3\columnwidth}
    \resizebox{\columnwidth}{!}{\begin{tikzpicture}
    \begin{axis}[
        width =\axisdefaultwidth,
        height= 180pt,
        legend style={font=\small},
        legend pos=south east,
        xlabel={\LARGE $\epsilon$},
        ylabel={\Large Time (s)},
        xtick=data,
        xticklabels={0.01, 0.025, 0.05, 0.075, 0.1},
        yticklabel style = {font=\Large},
        xticklabel style = {font=\Large},
        mark size=4pt,
        ymode=log]
        \addplot[mark=square, color=red] table [x=epsilon, y=milp-based, col sep=comma] {compas-eps.csv};
        \addplot[mark=o, color=blue] table [x=epsilon, y=k-level-based, col sep=comma] {compas-eps.csv};
        \legend{MILP-Based, $k$-level-based}
    \end{axis}
\end{tikzpicture}}
\subcaption{COMPAS, varying $\epsilon$}\label{hd-eps1}
\end{minipage}
\hspace{10pt}
\begin{minipage}[b]{0.3\columnwidth}
    \resizebox{\columnwidth}{!}{\begin{tikzpicture}
        \begin{axis}[
            width =\axisdefaultwidth,
            height= 180pt,
            legend style={font=\small},
            legend pos=south east,
            xlabel={\LARGE $\epsilon$},
            ylabel={\Large Time (s)},
            xtick=data,
            xticklabels={0.01, 0.025, 0.05, 0.075, 0.1},
            yticklabel style = {font=\Large},
            xticklabel style = {font=\Large},
            mark size=4pt,
            ymode=log,
            ymax=1.0]
            \addplot[mark=square, color=red] table [x=epsilon, y=milp-based, col sep=comma] {jee-eps.csv};
            \addplot[mark=o, color=blue] table [x=epsilon, y=k-level-based, col sep=comma] {jee-eps.csv};
            \legend{MILP-based, $k$-level-based}
        \end{axis}
    \end{tikzpicture}}
    \subcaption{IIT-JEE, varying $\epsilon$}\label{hd-eps2}
\end{minipage}
\caption{Runtime experimental results for multi-dimensional datasets ($3 \leq d \leq 6$).}
\end{figure}

\subparagraph{Baseline comparison.} For experiments in higher dimensions, the baseline algorithm failed to finish in almost all our experimental settings (with $d$ being the default and $n$ being $100\%$ of the dataset). The only exception was when $k=50$ and $\epsilon = 0.01$ for the ITT-JEE dataset (see Figure~\ref{md-baseline-compare}). In this case, our MILP-based and $k$-level-based algorithms achieved speedups of $3598$x and $4222$x respectively. For the 6-D dataset (COMPAS), we had to lower the value of $\epsilon$ ($\epsilon = 0.001$) to have the baseline algorithm finished within the time limit. As shown in Figure~\ref{md-baseline-compare}, our algorithms achieved greater speedups and were five orders of magnitude faster. Of course, these results aligns with the theoretical analysis, which predicts deteriorating performance of the baseline algorithm in higher dimensions. For a slightly larger $k$ value or $\epsilon$ value, the baseline algorithm failed to finish within the time limit due to the increased number of hyperplanes that needed to be inserted into the arrangement tree, while the run times of our two methods only  increased mildly. Due to the inefficiency of the baseline algorithm, for the rest of experiments, only run times of our two methods are going to be shown.

\subparagraph{Varying $\boldsymbol{k}$.} Figures~\ref{hd-k1} and \ref{hd-k2} shows the average run times of our two algorithms as $k$ varies. The MILP-based algorithm performed better than the $k$-level-based algorithm when $k$ was large, especially for the higher dimensional dataset (COMPAS). The $k$-level-based algorithm showed better performance for a small $k$ value and a lower dimensional dataset (IIT-JEE). It failed to finish within the time limit when the value of $k$ was large ($k \geq 70$) for the COMPAS dataset but succeeded in finishing within the time limit for the largest $k$ value ($k = 500$) in our experiments for the IIT-JEE dataset. Of course, these results are consistent with our theoretical analysis, as the structural complexity of the $k$-level increases with dimensionality. Unlike what we saw in the 2-D experiments, the run time of the multi-dimensional $k$-level-based algorithm is more dependent on the value of $k$. As each top-$k$ subset will generate $O(nk)$ linear programs to be solved, the run time is more sensitive to the $k$-level's structural complexity and the value of $k$ itself. On the other hand, the MILP-based algorithm also performed better for the lower dimensional dataset, even though the number of variables in the resulting mixed integer linear program was actually larger with the IIT-JEE being a larger dataset.

\subparagraph{Varying $\boldsymbol{\epsilon}$.} Figures~\ref{hd-eps1} and \ref{hd-eps2} show the average run times as $\epsilon$ varies. In both datasets, the run times of the $k$-level-based algorithm increased as $\epsilon$ increased. With a larger value of $\epsilon$, the breadth-first search will visit more cells, which results in longer run times. Moreover, the run time increased more dramatically for the higher dimensional dataset (COMPAS), which aligned with our theoretical analysis suggesting that the  structural complexity of the $k$-level grows with dimensionality. Meanwhile, the run times of the MILP-based algorithm also increased as $\epsilon$ increased, but experimental results show that the increase may become more mildly when the value of $\epsilon$ surpassed certain values for both datasets.

\subparagraph{Varying $\boldsymbol{n}$.} Figures~\ref{hd-n1} and \ref{hd-n2} show the average run times as $n$ varies. We controlled $n$ by randomly selecting a subset of data before preprocessing. In general, the performance of the $k$-level-based algorithm is more dependent on $n$, since a larger $n$ implies a larger number of $k$-level cells to be visited. However, we did see cases where the algorithm ran faster for a larger dataset. Of course, a larger dataset might not always end up with a larger number of $k$-level cells, especially when one only has to visit part of the $k$-level. Moreover, the run time scaled more dramatically for the higher dimensional dataset (COMPAS), which was aligned with our theoretical analysis. On the other hand, the run time of the MILP-based algorithm remained more stable across different numbers of $n$ in both datasets, even though the number of (indicator) variables increased as $n$ grew. This suggests that the stable performance of the MILP-based algorithm is likely due to its heuristic nature.

\subparagraph{Varying $\boldsymbol{d}$.} Figure~\ref{hd-d} shows the average run times as $d$ varies. We controlled $d$ by selecting a subset of scoring attributes of the COMPAS dataset after preprocessing. The run times of both algorithms increased significantly as the dimensionality increased. For the $k$-level-based algorithm, known theoretical results suggests that the structural complexity increases significantly as $d$ grows, which will greatly slow down the $k$-level-based algorithm. Moreover, the algorithm we used to solve linear programs actually runs in $O(d!n)$ time ($O(\cdot)$ hides the $d!$ factor in fixed dimensions), which also contributed to the run time increments with increasing dimensionality. For the MILP-based algorithm, even though the number of variables only increased by $1$ as $d$ increased by $1$, the run time still increased significantly. This suggests that the performance of the MILP-based algorithm is more sensitive to changes in $d$ (than in $n$). Notably, for $d=3$, the $k$-level-based algorithm was slower than the MILP-based algorithm, which was different from our observations with the IIT-JEE dataset (Figure~\ref{hd-k2}). We note that the 3-D datasets created by selecting three attributes contain many candidates with identical scoring attribute values, which exposed the inefficiency of the method for addressing tie-breaking challenge in our current $k$-level-based algorithm implementation. However, since this lower-dimensional dataset is derived from a higher-dimensional one, it may contain fewer scoring attributes than intended to distinguish each candidate. Therefore, the slowdown was more likely an artifact rather than a reflection of real-world scenarios. Nevertheless, our results are sufficient to show the performance trend of the $k$-level-based algorithm.

\subsubsection{Hardware utilization}\label{subsubsec:ablation}
This section presents results demonstrating the performance gains by our engineering optimizations, validating the effectiveness of our methods in addressing the hardware utilization challenges for the $k$-level-based algorithms, while also helping to distinguish between speedups attributable to algorithmic improvements and those arising from engineering optimizations~\cite{kriegel2017black}.

\begin{figure}[hbpt]
    \captionsetup[subfigure]{justification=centering}
    \centering
\begin{minipage}[b]{0.3\columnwidth}
    \resizebox{\columnwidth}{!}{\begin{tikzpicture}
        \begin{axis}[
            width =\axisdefaultwidth,
            height= 180pt,
            legend style={font=\small},
            legend pos=north west,
            xtick=data,
            symbolic x coords={COMPAS, IIT-JEE},
            xticklabels={\Large COMPAS\\\Large{($d=2$)}, \Large IIT-JEE\\\Large{($d=2$)}},
            x tick label style={
            align=center
            },
            bar width=20,
            enlarge x limits= 0.7,
            ylabel={\Large Speedup},
            xlabel={ },
            yticklabel style = {font=\Large},
            log origin=infty,
            ybar,
            ymode=log,
            nodes near coords,
            log ticks with fixed point,
            point meta=explicit symbolic,
            nodes near coords align={center},
            nodes near coords style={
            font=\large,
            black,
            yshift=6pt
            },
            ymax = 12,
            ytick={1, 2, 4, 8}]
            \addplot[darkgray, pattern=north east lines, pattern color=darkgray] coordinates {(COMPAS, 1.0)[1.0]  (IIT-JEE, 1.0)[1.0]};
            \addplot[teal, pattern=north west lines, pattern color=teal] coordinates {(COMPAS, 2.54) [2.54x] (IIT-JEE, 6.80) [6.80x]};
            \legend{Unoptimized, Optimized}
        \end{axis}
\end{tikzpicture}}
\subcaption{Memory access}\label{klevel-tree}
\end{minipage}
\hspace{10pt}
\begin{minipage}[b]{0.3\columnwidth}
    \resizebox{\columnwidth}{!}{\begin{tikzpicture}
        \begin{axis}[
            width =\axisdefaultwidth,
            height= 180pt,
            legend pos=south east,
            xlabel={\Large Number of Threads},
            ylabel={\Large Speedup},
            xmode=log,
            ymode=log,
            mark size=5pt,
            log ticks with fixed point,
            yticklabel style = {font=\Large},
            xticklabel style = {font=\Large},
            xtick={1, 4, 16, 64, 128},
            ytick={1, 4, 16, 64, 128}]
            \addplot[mark=triangle, color=OI6] table [x=threads, y=COMPAS, col sep=comma] {klevel-thread.csv};
            \addplot[mark=x, color=OI1] table [x=threads, y=JEE, col sep=comma] {klevel-thread.csv};
            \legend{\small{COMPAS}\normalsize{ ($\epsilon=0.025$)}, \small{IIT-JEE}\normalsize{ ($k=100$)}}
        \end{axis}
    \end{tikzpicture}}
    \subcaption{Parallelism}\label{klevel-thread}
\end{minipage}
\caption{Performance gains from better hardware utilization, measured by total run time.}
\end{figure}

\subparagraph{Memory access.} Figure~\ref{klevel-tree} shows the performance gains resulting from memory optimizations applied to the tournament tree implementation for the 2-D $k$-level-based algorithm. We compared the algorithm's overall performance using both an optimized and an unoptimized implementation, with the latter approximating a fully dynamic tournament tree while avoiding self-balancing by inserting new nodes into vacancies left by deletions. Consequently, the performance advantage of the optimized implementation over the unoptimized one primarily stems from memory optimizations. Our experimental results show notable speedups of $2.54$x and $6.80$x for the COMPAS and IIT-JEE datasets respectively, highlighting the practical performance benefits of our optimized implementation.

\subparagraph{Parallelism.} Figure~\ref{klevel-thread} shows the efficiency of our parallel implementation of the multi-dimensional $k$-level-based algorithm. For the COMPAS dataset, we set $k = 50$ and $\epsilon = 0.025$, while for the IIT-JEE dataset, we used $k = 100$ and $\epsilon = 0.05$. For our (stand-alone) sequential implementation, the average run times were $1503$s and $67$s respectively. With $64$ threads, our parallel implementation achieved speedups of $55$x and $47$x respectively, and with $128$ threads, the speedups increased to $102$x and $83$x. Our implementation demonstrated good scalability, effectively leveraging an increasing number of threads for heavy workloads without obvious degradation in parallel scaling performance, even up to 128 threads.

\subsection{Quality evaluation experiments}\label{subsec:expquality}
In outcome quality evaluation experiments, 6-D COMPAS dataset and 3-D IIT-JEE datasets were used, with $k=50$ and $k=100$ respectively. Recall that there are $50$ input weight vectors regardless of their initial fairness. The outcome quality of our two algorithms was evaluated using the following metrics:
\begin{itemize}
    \item \textbf{Found/Unfair ratio}: The ratio between the number of input weight vectors that are unfair and the number of corresponding fair vectors that are successfully found by the algorithm. (The results should be identical for both algorithms by the problem definition.)
    \item \textbf{$\boldsymbol{w}$ difference}: The average difference between the original unfair weight vector $w^o$ and the output fair weight vector $w^{\scriptscriptstyle f}$, measured by $||w^{\scriptscriptstyle f} - w^o||_1$, which itself can be interpreted as summing up all absolute weight differences. (Note that the value should not exceed $d\epsilon$.)
    \item \textbf{$\boldsymbol{\mathcal{G}_1}$ proportion}: The average proportion of the protected group ($\mathcal{G}_1$) candidates in the top-$k$ subset derived from the found fair weight vector. Ties were broken towards the $\mathcal{G}_1$ proportion in the entire dataset.
    \item \textbf{Utility loss}: By the principle in \cite{zehlike2022fairness}, the utility of a top-$k$ subset $\tau_k$ is computed by summing scores of all candidates in $\tau_k$ under the original weight vector $w^o$ as $U^o(\tau_k) = \sum_{c\in \tau_k} w^o \cdot p(c)$. The relative utility loss is then computed as $1 - \left. U^o(\tau^{\scriptscriptstyle f}_k) / U^o(\tau^o_k)\right.$, where $\tau^{\scriptscriptstyle f}_k$ and $\tau^o_k$ are top-$k$ subsets of $w^{\scriptscriptstyle f}$ and $w^o$, respectively. The average utility loss was reported. Ties were arbitrarily broken after applying the rule for $\mathcal{G}_1$ proportion.
\end{itemize}
These metrics were used to evaluate outcome quality under various fairness constraints and values of $\epsilon$. Note that the choice of fairness constraint often represents a societal and policy decision, but our algorithms maintain flexibility by accepting these constraints as input parameters.

\begin{table*}[tb!]
  \centering
  \caption{Outcome quality evaluation of our algorithms.}\label{tab:quality}
  \resizebox{\textwidth}{!}{
  \begin{tabular}{ |c|c|c|c|cc|cc|cc| }
    \hline
    \multirow{2}{*}{Dataset} & \multirow{2}{*}{$\epsilon$} & \multirow{2}{*}{\shortstack{Fairness \\ constraint}} & \multirow{2}{*}{\shortstack{Found/Unfair \\ ratio}} & \multicolumn{2}{c|}{$w$ difference} & \multicolumn{2}{c|}{$\mathcal{G}_1$ proportion} & \multicolumn{2}{c|}{Utility loss} \\
    & & & & $k$-level & MILP & $k$-level & MILP & $k$-level & MILP \\
    \hline
    \multirow{6}{*}{\shortstack{COMPAS\\($k=50$)}} & \multirow{3}{*}{0.05} & $[34\%, 66\%]$ & 19/39 & \textbf{0.149} & 0.211 & 66.0\% & \textbf{65.7\%} & \textbf{0.476\%} & 1.210\% \\
    & & $[40\%, 60\%]$ & 8/44 & \textbf{0.144} & 0.213 & 59.8\% & \textbf{59.5\%} & \textbf{0.449\%} & 1.328\% \\ 
    & & $[44\%, 56\%]$ & 6/47 & \textbf{0.155} & 0.202 & \textbf{55.3\%} & 55.6\% & \textbf{0.718\%} & 1.319\% \\ 
    \cline{2-10}
    & 0.025 & \multirow{3}{*}{$[40\%, 60\%]$} & 3/44 & \textbf{0.088} & 0.090 & 60.0\% & \textbf{59.3\%} & \textbf{0.121\%} & 0.142\% \\
    & 0.05 & & 8/44 & \textbf{0.144} & 0.213 & 59.8\% & \textbf{59.5\%} & \textbf{0.449\%} & 1.328\% \\ 
    & 0.075 & & 18/44 & \textbf{0.247} & 0.308 & 59.9\% & \textbf{59.7\%} & \textbf{2.416\%} & 3.187\% \\ 
    \hline
    \multirow{6}{*}{\shortstack{IIT-JEE\\($k=100)$}} & \multirow{3}{*}{0.05} & $[5\%, 45\%]$ & 7/22 & \textbf{0.055} & 0.089 & 5.00\% & 5.00\% & \textbf{0.033\%} & 0.067\% \\
    &  & $[7\%, 43\%]$ & 9/39 & \textbf{0.062} & 0.089 & 7.00\% & \textbf{7.22\%} & \textbf{0.025\%} & 0.043\% \\ 
    &  & $[9\%, 41\%]$ & 3/48 & \textbf{0.089} & 0.092 & 9.00\% & 9.00\% & 0.028\% & \textbf{0.027}\% \\
    \cline{2-10}
    & 0.025 & \multirow{3}{*}{$[9\%, 41\%]$} & 1/48 & 0.050 & 0.050 & 9.00\% & 9.00\% & 0.022\% & \textbf{0.020\%} \\
    & 0.05 & & 3/48 & \textbf{0.089} & 0.092 & 9.00\% & 9.00\% & 0.028\% & \textbf{0.027}\% \\ 
    & 0.075 & & 4/48 & \textbf{0.102} & 0.134 & 9.00\% & \textbf{9.25\%} & \textbf{0.036\%} & 0.096\% \\
    \hline
  \end{tabular}}
\end{table*}

Table~\ref{tab:quality} shows the experimental results for the two datasets. For the COMPAS datasets, as the fairness constraint became tighter, fewer input weight vectors satisfied it, and the algorithms were less likely to find a fair weight vector corresponding to a given unfair one. This outcome is expected since the problem becomes more restrictive with stricter constraints. Also, we note that increasing the value $\epsilon$ will increase the chance of finding a fair weight vector, which is intuitive since a larger weight vector space is going to be searched. In comparing the two algorithms, the $k$-level-based algorithm achieved better results in terms of the $w$ difference and utility loss. In fact, these two metrics are related: When $w^{\scriptscriptstyle f}$ is close to $w^o$, candidates with high scores under $w^{\scriptscriptstyle f}$ are also  likely to have high scores under $w^o$, resulting in a top-$k$ subset with higher utility and thus lower utility loss. Our $k$-level-based algorithm naturally favors ``nearby'' weight vectors due to its breadth-search approach with $w^o$ as the starting point, while the MILP-based algorithm does not directly leverage the knowledge of $w^o$. Nevertheless, both algorithms performed well, achieving small $w$ differences and utility losses. For the $\mathcal{G}_1$ proportion, both algorithms output weight vectors satisfying the fairness constraint, with the MILP-based algorithm being marginally better. As a side note, given the relation between $w$ difference and utility loss, we can also see that the range of fairness constraint only affects utility loss indirectly, through how close the nearest fair weight vector can be.

Similar performance patterns are observed on the IIT-JEE dataset. One thing to note is that finding fair weight vectors is generally more difficult for this dataset, requiring looser fairness constraints for the evaluation. Nevertheless, our algorithms still managed to increase the representation of protected group candidates in many cases.

\subsection{Observation summary and algorithm selection}\label{subsec:choice}
In this section, we summarize the key experimental observations (detailed analysis can be found above), especially for those that inform us the choice of algorithm in our two-pronged solution.

\subparagraph{2-D experiments.} In 2-D, our experiment results show that the $k$-level-based algorithm is the preferred method, with the following key observations:
\begin{itemize}
    \item Run times remained stable across $k$ for both datasets, suggesting that $k$-level structural complexity may have limited impact on run time in real-world 2-D scenarios.
    \item Varying $n$ experiments indicate that the $k$-level-based algorithm's advantage over the baseline may not due to the better worst-case complexity. Instead, it likely stems from the use of the kinetic priority queue (tournament tree) that directly reduces the number of pairwise line exchanges.
\end{itemize}
The first observation suggests that the algorithm may be practically efficient even for a relatively large $k$. Based on the second observation, our implementation of the $k$-level-based algorithm was purposely engineered to further improve the performance (Figure~\ref{klevel-tree}).

\subparagraph{Multi-dimensional experiments.} In higher dimensions, our experimental results show that both our algorithms significantly outperformed the baseline algorithm (Figure~\ref{md-baseline-compare}). They also produced high quality results (Table~\ref{tab:quality}). While the MILP-based algorithm being slightly worse in outcome quality, this can be mitigated by tuning $\epsilon$, making runtime efficiency the key factor in choosing between them. By our algorithm design principle (small $k$ vs. large $k$) outlined in Section~\ref{sec:design}, determining a threshold value of $k$ is crucial. However, the discussion in Section~\ref{subsec:small_k} does not apply here, since it only considers breaking worst-case lower bounds while our concern shifts to real-world performance. For real-world results, we have the following key observations:
\begin{itemize}
    \item The $k$-level-based algorithm outperformed the MILP-based algorithm for a sufficiently small $k$, but its run time grew more rapidly with $k$, unlike its stable performance in 2-D.
    \item Run times of both algorithms increased dramatically with dimensionality $d$, with the $k$-level-based algorithm increasing more rapidly, partly due to the LP algorithm used.
    \item The $k$-level-based algorithm performed better on the IIT-JEE dataset but worse on the 3-D COMPAS dataset with $k=50$ (Figures~\ref{hd-k2} and \ref{hd-d}), likely due to an artifact exposing an inefficiency in our implementation for tie-breaking.
\end{itemize}
Notably, our engineering efforts to leverage the multi-core system significantly enhance the performance of the $k$-level-based algorithm (Figure~\ref{klevel-thread}).

For the algorithm selection, the first two observations suggest that as $d$ increases, the threshold value of $k$ at which the MILP-based algorithm demonstrates superior performance is likely to decrease. However, the last observation suggests that performance comparisons between the two algorithms can be influenced by the underlying data distribution. Precisely defining a threshold value of $k$ for choosing between the two algorithms remains non-trivial, as the good performance of the MILP-based algorithm is inherently heuristic. This is also supported by the observation that its runtime is less sensitive to $n$ and the problem's NP-hardness. Nevertheless, our experimental results still indicate that this threshold is inversely correlated to $d$.

\section{Related work}
With decision-making increasingly being aided by automatic algorithms in many areas, such as hiring \cite{geyik2019fairness} and school admission \cite{peskun2007effectiveness}, ensuring fairness in this process has drawn increasing attention. Selecting the $k$-best items is an important task in decision-making. A typical method to do it is to assign a score to each item reflecting their fitness for a given target, and top-$k$ items are selected by  score. There is a substantial body of prior work \cite{zehlike2017fa, kleinberg2018selection, celis2018ranking, asudeh2019designing, yang2019balanced} exploring methods for achieving fair top-$k$ selection satisfying given proportional fairness constraints. See \cite{zehlike2022fairness} for a review.

In most previous work, only one attribute is used for scoring, or the score is pre-computed on multiple attributes with a given fixed scoring function. In many applications, however, one would like to select top-$k$ elements on multidimensional data where multiple attributes are relevant, and using an appropriate scoring function on these attributes to promote fairness could be a good choice. In \cite{asudeh2019designing}, the authors proposed algorithms to find an alternative linear scoring function if the given one does not produce results meeting the given fairness constraint. Since a linear scoring function can be presented as a weight vector, they aim at searching for a fair weight vector minimizing the angular distance. Their algorithms can work with various fairness models based on proportional fairness constraints, making them easily adaptable to our Fair Top-$k$ Selection problem. However, these algorithms do not scale efficiently with the size of dataset, and their run times increase dramatically with dimensionality (see Section~\ref{subsec:small_k} for more details). Besides applying proportional fairness constraints, the authors in \cite{liu2024fair} introduced a fairness metric, alpha-fairness, which is based on the differences between the proportions of protected group members in the top-$k$ subset and their proportions in the entire dataset. They further proposed methods for finding a fair weight vector optimizing this metric. However, their approach lacks a formal runtime analysis, offering only an outline that estimates a worst-case run time of $O(n^{2d - 2})$ but omits costs of some steps. Moreover, ties are arbitrarily broken in their methods, which may not be appropriate for fairness considerations (see Section~\ref{subsec:practical_klevel}).

Beside work on finding a fair linear scoring function, finding a linear scoring function that meets a specific need is also studied in other contexts, especially in reverse top-$k$ queries \cite{vlachou2010reverse, vlachou2013branch, chester2013indexing, gao2015answering, chen2023not}. Some reverse top-$k$ queries aim at finding linear scoring functions for a specific candidate being among top-$k$, with the candidate and the value of $k$ given as inputs. Most of these algorithms also do not scale well for multi-dimensional ($d\geq 3$) data. However, there are methods \cite{chen2023not} demonstrating good performances for multi-dimensional real-world datasets in practice, even though the worst case run time of the algorithm itself is not optimal.

The $k$-level-based algorithms presented in this work are based on the concept of $k$-level, a subject that has been extensively studied in computational geometry (see \cite{halperin2017arrangements}). There is a robust body of prior work on bounding the structural complexity of the $k$-level \cite{dey1998improved, toth2001point, sharir2001improved, agarwal1998levels,chiu2020average} and algorithms for constructing it \cite{edelsbrunner1986constructing, chan1999remarks, har2000taking, mulmuley1991levels, agarwal1998constructing, andrzejak1999optimization}. Unfortunately, empirical evaluations of these algorithms are limited, and some empirical results \cite{aharoni1999line, har2000taking} suggested that theoretically efficient algorithms may not perform well in practice.

\section{Conclusion}
We presented an integrated study on the problem of finding a fair linear scoring function for top-$k$ selection. Our work spans hardness analysis, algorithm design, practical engineering and empirical evaluations. We started with the hardness analysis for establishing the theoretical foundation and informing algorithm design (see Section~\ref{subsec:implications}). This was followed by algorithm design exploring both theoretically grounded and practically motivated opportunities, which lead to our two-pronged solution: A $k$-level-based algorithm for small $k$ and a MILP-based algorithm for large $k$. Practical engineering then addressed implementation challenges arising from the theory-practice gap and fairness considerations inherent to the problem (see Section~\ref{subsec:practical_klevel}). Finally, empirical evaluation explored scenarios where worst cases behavior does not manifest and identified areas critical to real-world performance. (See Figure~\ref{fig:structure} for a more detailed roadmap.) Our result is a practically efficient solution that is significantly faster than SOTA, accompanied by many insights which we believe can inform future studies.

As for the future direction, although our $k$-level-based algorithms are significantly faster for small $k$, they still do not scale efficiently for any $d > 3$, with the best run time being $O(n^{\lfloor d/2 \rfloor}k^{\lceil d/2 \rceil})$. It would be interesting to see if there exist better algorithms with run times such as $O(n + k^{d-1})$. Such an algorithm would not violate the theoretical limits while offering significantly better scalability for small $k$. Moreover, it would be interesting to see how to make such an algorithm work in practice. On the other hand, it is well known that a MILP solver applies many heuristics that utilize the structure of the given program for the speedup, which results in the better real-world performance for our problem. It would be interesting to see how far one can go using methods that explicitly utilize the structure of the dataset itself, particularly since our hardness analysis shifts the focus to real-world performance for large $k$. In this context, understanding the typical structure of an input dataset could be valuable.

\bibliographystyle{plainurl}
\bibliography{ref}

\end{document}